\documentclass[msom,nonblindrev]{informs3_hide}

\OneAndAHalfSpacedXI % default for MS

\newcommand\linespace{1.6}

\usepackage[english]{babel}
\usepackage[autostyle, english = american]{csquotes}
\MakeOuterQuote{"}

\usepackage[dvipsnames]{xcolor}
\usepackage{color}    
\usepackage{color-edits}
\usepackage{siunitx}
\definecolor{darkgreen}{rgb}{0.0, 0.5, 0.0}  % Custom dark green
\definecolor{darkmagenta}{rgb}{0.5, 0.0, 0.5}  % Custom dark magenta
\definecolor{darkblue}{rgb}{0.0, 0.0, 0.5}   % Custom dark blue
\definecolor{darkred}{rgb}{0.5, 0.0, 0.0}    % Custom dark red

\definecolor{myBlue}{RGB}{30,144,255}    % DodgerBlue: bright, saturated blue
\definecolor{myOrange}{RGB}{255,165,0}     % Orange
\definecolor{myGreen}{RGB}{34,139,34}      % ForestGreen (or try a brighter green like lime: {50,205,50})
\definecolor{myMagenta}{RGB}{255,0,255}    % Magenta

\usepackage{bbm, xspace}
\usepackage[hypertexnames=false]{hyperref}
\usepackage{cleveref}
\crefname{subsection}{subsection}{subsections}
\usepackage{enumitem}
\usepackage[normalem]{ulem}

\usepackage{caption}
\usepackage{subcaption}

\usepackage{pgfplots}
\usepgfplotslibrary{groupplots}
\pgfplotsset{compat=1.18}
\usetikzlibrary{decorations.markings, shapes.geometric}
\pgfplotsset{
    cycle list={
        {myBlue, mark=*, dashed, thick, mark options={solid, fill=myBlue}},
        {myOrange, mark=square*, dashed, thick, mark options={solid, fill=myOrange}},
        {myGreen, mark=triangle*, dashed, thick, mark options={solid, fill=myGreen}},
        {myMagenta, mark=diamond*, dashed, thick, mark options={solid, fill=myMagenta}}
    }
}

\usepackage{pgfplotstable} % For handling tables from CSV files
\usepackage{booktabs}      % For professional-looking tables

\newcommand{\eps}{\varepsilon}

\newcommand{\bE}{\mathbb{E}}
\newcommand{\bR}{\mathbb{R}}
\newcommand{\bZ}{\mathbb{Z}}

\newcommand{\OPT}{\mathsf{OPT}}

\addauthor{Will}{red}

\usepackage{natbib,bbm,multirow,multicol}
 \bibpunct[, ]{(}{)}{,}{a}{}{,}%
 \def\bibfont{\small}%
\TheoremsNumberedThrough     % Preferred (Theorem 1, Lemma 1, Theorem 2)
\ECRepeatTheorems

\EquationsNumberedThrough    % Default: (1), (2), ...
\usepackage{tabstackengine}
\usepackage[nopar]{lipsum}

\usepackage{algorithm}
\usepackage{algpseudocode}

\newcommand{\PP}{\mathbb{P}}
\newcommand{\E}{\mathbb{E}}

\newcommand{\R}{\mathbb{R}}

\newcommand{\Ind}{\mathbbm{1}}

\newcommand{\N}{\mathbb{N}}

\newcommand{\ONL}{\mathsf{ONL}}

\newcommand{\MYO}{\mathsf{MYO}}
\newcommand{\OFF}{\mathsf{OFF}}
\newcommand{\OFFM}{\mathsf{OFF}_{\mathsf{multi}}}
\newcommand{\FLU}{\mathsf{FLU}}

\newcommand{\CH}{\mathrm{CH}}

\newcommand{\HH}{\mathrm{H}}
\newcommand{\LL}{\mathrm{L}}

\newcommand{\hOFF}{\widehat{\OFF}}

\newcommand{\flxi}{\lfloor x_i \rfloor}
\newcommand{\flxil}{\lfloor x_{i\ell} \rfloor}
\mathchardef\mhyphen="2D % Define a "math hyphen"
\newcommand{\DHTI}{\mathsf{DH\mhyphen TI}}
\newcommand{\RHTI}{\mathsf{RH\mhyphen TI}}
\newcommand{\ROSI}{\mathsf{RO\mhyphen SI}}

\newcommand{\X}{\mathcal{X}}
\newcommand{\XM}{\mathcal{X}_{\mathsf{multi}}}

\newcommand{\samplePathOld}{{}}

\begin{document}
\linespread{\linespace}\selectfont{}

%%%%%%%%%%%%%%%%

% Outcomment only when entries are known. Otherwise leave as is and
%   default values will be used.
%\setcounter{page}{1}
%\VOLUME{00}%
%\NO{0}%
%\MONTH{Xxxxx}% (month or a similar seasonal id)
%\YEAR{0000}% e.g., 2005
%\FIRSTPAGE{000}%
%\LASTPAGE{000}%
%\SHORTYEAR{00}% shortened year (two-digit)
%\ISSUE{0000} %
%\LONGFIRSTPAGE{0001} %
%\DOI{10.1287/xxxx.0000.0000}%

%\RUNAUTHOR{}

%\RUNTITLE{}

\TITLE{Optimizing Inventory Placement for a Downstream Online Matching Problem}

\ARTICLEAUTHORS{
\AUTHOR{Boris Epstein}
\AFF{Graduate School of Business, Columbia University, New York, NY 10027, \EMAIL{bepstein25@gsb.columbia.edu}}

\AUTHOR{Will Ma}
\AFF{Graduate School of Business and Data Science Institute, Columbia University, New York, NY 10027, \EMAIL{wm2428@gsb.columbia.edu}}
}

\ABSTRACT{

% \textbf{Problem definition:}
We study the inventory placement problem of splitting $Q$ units of a single item across warehouses in advance of a downstream online matching problem that represents the dynamic fulfillment decisions of an e-commerce retailer.
This is a challenging problem both theoretically, due to the computational complexity of the downstream matching problem, and practically, as the fulfillment team continuously updates its algorithm while the placement team lacks direct evaluation of placement decisions.
% \textbf{Methodology/results:}

We compare the performance of three placement procedures based on optimizing surrogate functions that have been studied and applied: Offline, Myopic, and Fluid placement.
On the theory side, we show that optimizing inventory placement for the Offline surrogate leads to an $\alpha (1-(1-1/d)^d)$-approximation for the joint placement and fulfillment problem under any demand model that admits an $\alpha$-competitive fulfillment policy. We assume $d$ is an upper bound on how many warehouses can serve any demand location.
The crux of our theoretical contribution is to use randomized rounding to derive a tight $(1-(1-1/d)^d)$-approximation for the integer programming problem of optimizing the Offline surrogate. We further show how to extend this result to a multi-SKU setting, improving upon the best known approximation of $1/2$.
% We show that this approximation ratio cannot be obtained by other approaches using the Fluid surrogate or submodular optimization, and also show that it is tight.
We use statistical learning to show that rounding after optimizing a sample-average Offline surrogate, which is necessary due to the exponentially-sized support, indeed has vanishing loss. 
On the experimental side, we evaluate how different combinations of placement and fulfillment procedures perform on a wide array of synthetic instances.
%we extract real-world sequences of customer orders from publicly-available JD.com data and evaluate different combinations of placement and fulfillment procedures.
% \bedit{This evaluation differs from theory in that it is average-case (not worst-case), does not make independence assumptions about demand arrivals, and does not assume training/test data are drawn from the same distribution.}
When coupled with a good fulfillment procedure, optimizing the Offline surrogate performs best even compared to computationally-intensive simulation procedures, corroborating our theory.

% \textbf{Managerial implications:}
Theoretical guarantees and extensive numerics suggest that the placement team should optimize the (optimistic) Offline surrogate, assuming the fulfillment team has a good algorithm.  Otherwise, the placement team could optimize the (pessimistic) Myopic surrogate instead.
}

%\KEYWORDS{}

%\HISTORY{}

\maketitle
%%%%%%%%%%%%%%%%%%%%%%%%%%%%%%%%%%%%%%%%%%%%%%%%%%%%%%%%%%%%%%%%%%%%%%

% Paragraph describing practical scope: e-commerce, different stages

% Paragraph justifying the importance of simplifying assumptions. Realistically it’s always separate teams deciding this, they don’t solve the problem jointly. Fulfillment team takes placement and instance as input, outputs a fulfillment policy. Placement team takes an instance and must output a placement, . \wnote{Important point to mention: the downstream problem is computationally challenging, which is the main reason the upstream team needs to make simplifying assumptions.  Otherwise, one could simply prescribe the jointly optimal solution.}

% Paragraph describing the joint placement and fulfillment problem

% Here mention OFF and MYO and FLU with different papers backing each one,

% Subsection: Empirical findings

% \bnote{Multiple items setting:
% \begin{itemize}
%     \item $i\in[n]$ warehouses
%     \item $j\in[m]$ demand nodes
%     \item $k\in[s]$ skus
%     \item $\omega \in \Omega$ stochastic scenarios
%     \item warehouse capacities $C_i$
%     \item quantities to stock $Q_k$
% \end{itemize}
% LP:
% \begin{align*}
%     \max_{x,y\geq 0} \quad &\sum_{\omega, i,j,k} p^\omega y_{ijk}^\omega r_{ijk}\\
%     s.t. \quad &\sum_{i} x_{ik} \leq Q_k &\forall k\\
%     & \sum_{k} x_{ik} \leq C_i &\forall i\\
%     & \sum_{i} y_{ijk}^\omega \leq D_{js}^\omega &\forall j,k,\omega\\
%     & \sum_{j} y_{ijk}^\omega \leq x_{ik} & \forall i,k,\omega
% \end{align*}

% }

% \Willcomment{You will add this in a new section?}

% \bnote{Good question. }

\section{Introduction}

\linespread{\linespace}\selectfont{}

The supply chain of an e-commerce retailer is a large and complicated system whose overall operation results from an extensive stream of interdependent decisions.
These decisions range all the way from long-term strategic decisions, including network design and sourcing, to mid-term tactical decisions, including inventory replenishment and placement, to real-time operational decisions, including order fulfillment and last-mile delivery \citep{chen2021item}. 
The optimality of upstream decisions depends on how downstream decisions will be made, and vice-versa.
However, jointly optimizing different stages is computationally and practically difficult \citep{vanryzin2024consensus,maggiar2024consensus}.
%, and moreover, the sheer size of modern tech companies inevitably leads to a hierarchical structure where a separate team controlls each.
Therefore, a common approach for making upstream decisions is to abstractly model the downstream dynamics, or make simplifying assumptions on how downstream decisions will be made.
%\footnote{The ideal way for teams to coordinate, and communicate updates in their algorithm so that other teams can adjust accordingly, is a fascinating problem for both theory and practice.}

In this paper, we study the problem of an e-commerce retailer faced with making \emph{inventory placement} decisions followed by on-the-fly \emph{fulfillment} decisions. The e-commerce retailer manages a network of warehouses that can deliver goods to several different last-mile delivery hubs. Before orders arrive, the e-commerce retailer must decide how to split an incoming purchase of inventory units across the multiple warehouses it manages. This corresponds to the {inventory placement problem}. Afterward, the e-commerce retailer faces a stream of customer orders during a time horizon, which is random and therefore not known in advance. Whenever an order is made, the e-commerce retailer must decide which warehouse to deliver from, depleting a unit of inventory and collecting a reward. This sequential decision-making problem corresponds to the {fulfillment problem}. 
The goal of the e-commerce retailer is to maximize the expected reward collected during the time horizon by deciding how to place inventory and how to fulfill demand. These two decisions are tightly linked together. On one hand, fulfillment decisions depend on the current inventory levels at each warehouse, which in turn depend on the initial inventory placement. On the other hand, the optimal inventory placement depends on how fulfillment decisions are going to be made. Despite this connection, only recently has there been increased attention on how these decisions interact and how to make them in a joint or coordinated manner \citep{govindarajan2021joint,bai2022coordinated,chen2022approximation,arlotto2023online,jasin2024inventory}.

As an isolated problem, online fulfillment has been well-studied by the operations research and computer science communities \citep{xu2009benefits,acimovic2015making,jasin2015lp}. The main tool of analysis for this problem is modeling it as an online bipartite matching problem {with stochastic inputs \citep{feldman2009online,alaei2012online,brubach2020online}}. These models take the initial inventories at different warehouses as input, and the policies developed aim to obtain a reward as close as possible to the best-achievable reward with that inventory.
{On the other hand, the inventory placement problem has been less studied.  Here, a fixed quantity of an item that has already been purchased is to be distributed among a set of warehouses.}
A fundamental difference between the placement and fulfillment problems is that there are no downstream decisions after fulfillment is done, whereas the effectiveness of an inventory placement directly depends on the fulfillment policy being deployed. For a fixed inventory placement, finding the optimal fulfillment policy can be computationally challenging \citep{papadimitriou2021online}, {which means that even evaluating the potential value of a proposed placement is hard.}

To overcome this difficulty, an approach taken by the literature and also in practice is to (exactly or approximately) optimize a surrogate function that approximates the value of the optimal fulfillment policy. The following surrogates and corresponding placement procedures have been considered.
\begin{enumerate}
\item \textbf{Fluid Placement}: optimizes inventory placement for a fluid relaxation of the problem, in which the total demand of each node is assumed to be deterministic and equal to its expected value. This procedure only requires knowing the first moments of demand distributions and is also the least computationally expensive.  It has been used to prove some approximation guarantees for the joint placement and fulfillment problem \citep{bai2022joint,chen2022approximation}.
\item \textbf{Offline Placement}: optimizes the expected reward obtained by a fulfillment policy that knows the total demand for each node in the given random scenario in advance and thus makes hindsight optimal fulfillment decisions. \citet{govindarajan2021joint,devalve2023understanding} advocate for this procedure based on their experiments.
% , though without theoretical justification.
{This procedure requires the distributional knowledge of total demands and the ability to enumerate over it, which can be difficult.  Therefore, it is often approximated by the sampling of demand scenarios.}
\item \textbf{Myopic Placement}: makes the simplifying assumption that the downstream fulfillment decisions will be made according to a greedy policy that maximizes immediate rewards, ignoring the opportunity cost of depleting a unit of inventory at a given warehouse. This is an approach present in the lines of work of \citet{acimovic2017mitigating,chen2017large}, which presents it as a natural first step.
% , albeit without apparent empirical or theoretical justification.
Unlike the first two procedures, the surrogate optimized here is rather pessimistic about the sophistication of the downstream fulfillment team.
% , making it a less suitable object for theoretical analysis aiming to obtain approximation guarantees.
The output of myopic fulfillment depends on the order in which demand arrives, so this information has to be incorporated when optimizing inventory placement for this surrogate, which is usually done through simulation.
\end{enumerate}
Despite the aforementioned work, there is no clear consensus on which placement procedure is best.
Fluid Placement comes with theoretical guarantees, but is shown to perform substantially worse than Offline Placement in real-world experiments \citep{devalve2023understanding}.
Offline Placement, on the other hand comes with no theoretical guarantees, and the number of samples required for it to prescribe a good decision is also unclear.
Finally, Myopic Placement is the most realistic in terms of not expecting the fulfillment team to foretell demand in advance, but the speed and stability of simulation procedures is questionable.

In this paper, we take a deep dive into understanding how these three common approaches for optimizing inventory placement compare. Our contributions come both from deriving theoretical results and from {numerical simulations.}
%conducting experiments based on publicly-available real-world data.
On the theoretical side, we show that when coupled with an appropriate, high-quality fulfillment policy, Offline Placement offers constant-factor guarantees for the joint placement and fulfillment problem.
% \Willdelete{To the best of our knowledge, we are the first work to establish this kind of guarantee using Offline Placement as the inventory placement procedure.}
Moreover, our guarantee is strictly better than the best guarantees obtained by using Fluid Placement for the inventory placement, complementing the empirical findings of \citet{devalve2023understanding}{, and it holds on a strictly larger class of demand models}. This result formalizes the notion that Offline Placement, a procedure that can be interpreted as being optimistic about the quality of the downstream fulfillment decisions, is desirable when the fulfillment policy being deployed is of high quality. 
{Our numerical simulations corroborate our theoretical finding.}
%\Willcomment{TODO decide if we still want to mention these details}

%\bedit{On the numerical side, we {extensively and thoroughly} compare the performance of different placement procedures on a wide variety of instances. Our results suggest that Offline Placement robustly outperforms Myopic and Fluid placements when coupled with a high-quality fulfillment policy, corroborating our theoretical findings.}
% They also suggest that Myopic Placement has superior performance to Fluid Placement, supporting the approach of \citet{acimovic2017mitigating}.}

% On the experimental side, we {extensively and thoroughly} compare the performance of different placement procedures on the JD.com dataset from the 2020 MSOM Data-driven Research Challenge \citep{shen2020jd}. Our results suggest that Myopic Placement has superior performance to Fluid Placement, supporting the approach of \citet{acimovic2017mitigating}.  The comparison between Offline and Myopic Placement is a closer one however; which one outperforms the other is dependent on the fulfillment policy being used and the instance parameters. Our results suggest that Offline Placement has a more robust performance when coupled with a high-quality fulfillment policy, which corroborates our theoretical finding.

\subsection{Theoretical Results and Techniques \label{subsec:theo_contributions}}
\linespread{\linespace}\selectfont{}

The core of our theoretical contribution is an approximation algorithm for inventory placement optimization under the Offline surrogate, which is developed in \textbf{\Cref{sec:techinical}}. This is a fundamental integer programming problem related to matching, defined
% that is surprisingly non-trivial and, to the best of our knowledge, previously unsolved. The problem is
as follows.  We are given a bipartite graph with warehouses $i\in[n]:=\{1,\ldots,n\}$, demand nodes $j\in[m]$, and rewards $r_{ij}\ge0$ such that $|\{i:r_{ij}>0\}|\le d$ for all $j$. Here, $d$ measures fulfillment flexibility, which can also be interpreted as an upper bound on the number of warehouses that receive a positive reward for serving any given demand location.  {(Our results are relevant even if $d=\infty$, representing full flexibility.)}  A demand vector $D=(D_1,\ldots,D_m)$ is drawn from an arbitrary distribution over $\bZ_{\ge0}^m$.  The problem is to decide an inventory placement $x=(x_1,\ldots,x_n)\in\bZ_{\ge0}^n$, subject to a constraint $x_1+\cdots+x_n=Q$ where the total inventory $Q$ has been decided already, to maximize the expected value of
\begin{align}
\OFF(x,D)= \max_{y\geq0} \quad & \sum_{i\in[n]}\sum_{j\in[m]} y_{ij} r_{ij}\nonumber\\
\mathrm{s.t.}\quad & \sum_{i\in [n]} y_{ij} \leq D_j \quad \forall j\in[m],\label{const:demandSide} \\
&\sum_{j\in[m]} y_{ij}\leq x_i \quad \forall i\in[n],\label{const:supplySide}
\end{align}
over the randomly-drawn demand vector $D$.  Assuming for now that the distribution of $D$ has small support, the fractional relaxation of the placement problem,
\begin{align} \label{eqn:relaxationIntro}
\max_{x\in\bR^m_{\ge0},\sum_i x_i=Q}\bE_{D}[\OFF(x,D)],    
\end{align}
can be formulated as an LP by enumerating the possible realizations of $D$.
% It remains to show how to round a fractional solution $x$ into $\bZ_{\ge0}^n$ while incurring minimal loss.
It is important to note that the optimal solution of~\eqref{eqn:relaxationIntro} can indeed be fractional even though the bipartite matching LP for any fixed $D$ is integral.  In fact, we show that the value of~\eqref{eqn:relaxationIntro} can shrink by a factor of $1-(1-1/d)^d$ when $x$ is constrained to lie in $\bZ^n_{\ge0}$ (\Cref{lem:tight_example} in \textbf{\Cref{subsec:rounding}}). Our main result is to show that this factor is in fact tight, i.e.\ any fractional placement $x$ can be rounded into an integer placement $R(x)=(R_1(x),\ldots,R_n(x))\in\bZ_{\ge0}^n$ such that $\bE_{D}[\OFF(R(x),D)]\ge(1-(1-1/d)^d)\bE_{D}[\OFF(x,D)]$ (\Cref{thm:rand_rounding} in \textbf{\Cref{subsec:rounding}}).
We develop a randomized rounding for this problem based on applying the fundamental rounding procedure of \citet{gandhi2006dependent} in two iterations. In the first iteration, we round the inventory placements obtained in~\eqref{eqn:relaxationIntro} and in the second iteration we round the LP variables defining $\OFF(x,D)$ for every possible outcome of the demand vector $D$. This second iteration rounding must depend on the outcome of the first iteration, which requires a new rounding subroutine that we describe in \textbf{\Cref{subsec:rounding}}.

Our use of the Offline relaxation assumes that we can enumerate over the support of $D$, whereas in general, the support of $D$ has exponential size. To fix this, we show that given $K$ independent and identically distributed (IID) samples, we can still obtain an approximation ratio of $1-(1-1/d)^d-O(Q\sqrt{(n\log n)/K})$ when Offline Placement solves a sample-average approximation of~\eqref{eqn:relaxationIntro} using the $K$ samples, whose size is polynomial in $K$. We use Rademacher complexity and a vector contraction inequality \citep{maurer2016vector} to show that the sample-average approximation provides a uniform approximation of $\bE_D[\OFF(x,D)]$ over the continuous feasible region $\{x\in\bR^m_{\ge0}:\sum_i x_i=Q\}$ (\Cref{thm:stat_learning} in \textbf{\Cref{subsec:statLearning}}), which completes the argument.  Although this argument is simple in hindsight, we remark that a typical analysis of sample-average approximation for optimization problems \citep[e.g.]{shapiro2005complexity} does not go through uniform convergence. Consequently, we would lose an additional factor of 2 in the approximation ratio instead of losing a factor that is vanishing as $K\to\infty$ if the optimization problem is not solved to optimality.
% \Willcomment{Maybe just don't explain here as to avoid causing confusion on whether people have previously used submodular to analyze Offline.}
% (In particular, if we were to use submodular optimization to obtain a $1-1/e$-approximation for the SAA version of $\max_{x\in\bZ^m_{\ge0},\sum_i x_i=Q}\bE_{D}[\OFF(x,D)]$ as in \cite{bai2022joint}, and use the sample-average approximation analysis from \citet{shapiro2005complexity}, we would only obtain a $(1-1/e)/2$-approximation for the true problem, even if $K\to\infty$.)

In \textbf{\Cref{sec:joinproblem}} we apply these results to the joint placement and fulfillment problem under the assumption that {our demand model admits an $\alpha$-competitive online fulfillment policy. This means that the setting is such that, no matter what initial inventories we have, we can efficiently compute a fulfillment policy that collects on average at least $\alpha$ times what we would collect if we knew the stream of customer orders in advance and made hindsight optimal fulfillment decisions. We show that, when paired with this $\alpha$-competitive fulfillment policy, Offline Placement collects a reward that is at least $\alpha (1-(1-1/d)^d)$ times what the optimal joint placement and fulfillment procedures would collect}, where $d$ is the maximum number of warehouses that can fulfill a single demand node (\Cref{thm:main} in \textbf{\Cref{subsec:mainResult}}).
% (\Cref{thm:main}, \Cref{subsec:mainResult}).
%To our knowledge, this is the first result showing that Offline Placement achieves a constant-factor guarantee for the joint placement and fulfillment problem.
Our general treatment of demand models allows us to recover existing and obtain new results in relevant settings. For example, the Temporal Independence model (classic revenue management and online matching model with deterministic, known time horizon and independent arrivals; e.g., \citet{alaei2012online}) allows a $1/2$-competitive fulfillment policy. In this setting, we recover the $(1-1/e)/2$-approximation implied by \citet{bai2022joint} and strictly improve it for small values of $d$.
%we strictly improve the $(1-1/e)/2$-approximation implied by \citet{bai2022joint} {(their guarantee is stated as 1/4 and holds in a generalized setting; it improves to $(1-1/e)/2$ in the setting stated here)}.
To obtain their result, \citet{bai2022joint} uses Fluid Placement, so our result also provides theoretical justification for why Offline appears to outperform Fluid Placement in the experiments of \citet{devalve2023understanding}.
%As $d\to\infty$, our guarantee decreases to approach their guarantee of $(1-1/e)/2$, which is obtained by using submodular optimization to approximately solve Fluid Placement within a factor of $1-1/e$. 
%However, the submodular optimization approach cannot improve (see \textbf{\Cref{sec:greedySucks}}) as $d$ gets small, which is often the case in practice---$d=2$ in the celebrated long-chain network \citep{jordan1995principles} or the JD.com network \citep{shen2020jd}.
%This is why we take a randomized rounding approach instead, and moreover use it to analyze Offline Placement since we cannot use integrality properties about demand in Fluid Placement.
Another example setting is the Spatial Independence model of \citet{aouad2022nonparametric}. Here, the authors establish a $1/2$-competitive fulfillment policy, so our result implies a $(1-(1-1/d)^d)/2$-approximation in this setting.
Importantly, the 1/2-competitive fulfillment policy of \citet{aouad2022nonparametric} only works against the weaker Offline surrogate, so previous approaches would not have obtained any constant-factor approximation ratios at all using the Fluid surrogate.
% Moreover, they show that the fluid relaxation is too loose in this setting to yield constant-factor competitive ratios for online algorithms, highlighting another benefit of using Offline Placement over Fluid Placement.
% Another benefit of using Offline Fulfillment over Fluid Placement is that in some arrival models of online matching, the fluid relaxation is too loose to yield constant-factor competitive ratios for online algorithms, as is the case in the Spatial Independence model by \citep[Proposition 1]{aouad2022nonparametric}.

In \textbf{\Cref{sec:multisku}} we show how our randomized rounding result for the Offline surrogate extends to the multi-SKU setting from \citet{bai2022joint}. In this setting, the e-tailer stores and faces demand for $s$ different SKUs that must share capacities at the different warehouses. The key observation for extending our result to this setting is that after using the algorithm of \citet{gandhi2006dependent} in the first iteration to round the inventory placement, the fulfillment decisions can be separated across SKUs. Therefore, we can use the same second iteration rounding for all pairs of SKUs and demand vectors for the particular SKU. That way, our rounding guarantee of $1-(1-1/d)^d> 1-1/e$ still holds in this multi-SKU setting.
{Importantly, any attempt to prove this result using submodularity would now have to consider the intersection of two partition matroids, for which the best-known approximation factor is only 1/2 {\citep{lee2010maximizing}}, worse than our guarantee even if $d=\infty$.}

\subsection{Numerical Results From Simulations}
\linespread{\linespace}\selectfont{}

In \textbf{\Cref{sec:synthexp}}, we complement our theoretical results with simulations. We evaluate how different placement and fulfillment procedures perform when deployed together in a wide array of instances with different network structures, demand models, and load factors (ratio between expected demand and number of inventory units).
The main difference between our theoretical results and our simulations is that we drift from the adversarial nature.
% experiments using the JD.com data available from the 2020 MSOM Data-driven Research Challenge \citep{shen2020jd}. {These experiments are inspired by the ones conducted in \citet{devalve2023understanding}.} We use this dataset to evaluate how different placement and fulfillment procedures perform when deployed together in this particular JD.com network structure. The main feature we extract from the data, and what makes these experiments different from our theoretical settings, is the exact arrival sequences of demand. 
{Indeed, whereas our theoretical results provide worst-case guarantees for all possible network structures and demand distributions for a given demand model, our simulations consider particular instances motivated by the fulfillment literature.}
%The theoretical settings also impose assumptions on the stochastic demand process, which is not the case in the dataset. Moreover, the theoretical results assume that the demand distribution is known and we can generate IID samples. In our experiments, we use past demand data as samples for our placement and fulfillment procedures and evaluate them out-of-sample on future demand, where there may have been distributional shifts in reality. %\wnote{Amazing paragraph!}

We evaluate four different placement procedures: Offline Placement, Myopic Placement, and two versions of Fluid Placement. The fulfillment procedures evaluated are Myopic Fulfillment and several variants of linear programming-based policies that use the shadow prices of their constraints to approximate and incorporate the opportunity costs of depleting inventory from specific warehouses. The variants depend on the linear program being solved (fluid or offline) and whether they include re-solving or not.

Consistent with \citet{devalve2023understanding}, we find that the fulfillment policy that uses the shadow prices of the offline linear program and re-solves periodically has the best performance. Moreover, when this fulfillment policy is deployed, Offline Placement achieves the best performance among the four placement procedures we benchmark. This result is robust to 
varying network structures, demand models, and load factors.
This again backs the notion that deploying Offline Placement leads to good performance, given that the fulfillment policy is of high quality. We also find that Myopic Placement outperforms both versions of Fluid Placement in most of our instances, suggesting that using a placement procedure that takes into account the uncertainty of the demand (Offline Placement or Myopic Placement) is consistently better than using deterministic approximations that use average demand as input (Fluid Placement).
% can be a desirable placement procedure to deploy if edge rewards are similar among each other, and thus the main cause for losing reward is lost sales; {or of course, if Myopic Fulfillment will indeed be the fulfillment policy.}
% {Overall,} our experiments suggest that using a placement procedure that takes into account the uncertainty of the demand (Offline Placement or Myopic Placement) is consistently better than using deterministic approximations that use average demand as input (Fluid Placement or Proportional Placement){---and this is despite the fact that the more sophisticated Offline and Myopic Placement procedures are more likely to overfit to the training data.}

\Cref{tab:intro_summary} presents a summary of how the three placement procedures of interest compare and how our results contribute to this understanding.

\newcommand{\offwidth}{.33}
\newcommand{\myowidth}{.20}
\newcommand{\fluwidth}{.27}
\begin{table}[]
\linespread{1.2}\selectfont{}
\small
    \centering

\begin{tabular}{|p{.12\textwidth}|p{\offwidth\textwidth}|p{\myowidth\textwidth}|p{\fluwidth\textwidth}|}
\hline
\textbf{Placement Procedure}     & \textbf{Offline Placement $(\OFF)$}                                                                                                   & \textbf{Myopic Placement $(\MYO)$}                            & \textbf{Fluid Placement $(\FLU)$}                                                                            \\ \hline
Information Required    & Medium (all moments of aggregate demand for each node)                                                   & High (exact demand arrival sequence) & Low (first moment of aggregate demand for each node)                              \\ \hline
Computation Time        & Requires sampling                                                                                        & Requires simulation                  & Fastest (still no poly-time algorithm)                                            \\ \hline
Surrogate Approximation & \begin{tabular}[c]{@{}p{\offwidth\textwidth}@{}}$1-(1-1/d)^d$ $\boldsymbol{(*)}$ \\ (LP and rounding)  \end{tabular}                                                                                     &             no known guarantees                         & \begin{tabular}[c]{@{}p{\fluwidth\textwidth}@{}} $1-1/e$     \citep{bai2022joint}\\ (Submodular optimization)\end{tabular}                                                             \\ \hline
Overall Approximation   & \begin{tabular}[c]{@{}p{\offwidth\textwidth}@{}}$\alpha (1-(1-1/d)^d)$ $\boldsymbol{(*)}$\\ (Wtih $\alpha$-approximate fulfillment policy against Offline)\end{tabular}                   &                     no known guarantees                 & \begin{tabular}[c]{@{}p{\fluwidth\textwidth}@{}}$\alpha (1-1/e)$ \citep{bai2022joint}\\ (Wtih $\alpha$-approximate fulfillment policy against Fluid)\end{tabular} \\ \hline
Empirical Justification & \begin{tabular}[c]{@{}p{\offwidth\textwidth}@{}}Beats $\FLU$ \citep{devalve2023understanding}\\ Beats $\MYO$ given a good fulfillment policy  $\boldsymbol{(*)}$  \end{tabular} & Beats $\FLU$  $\boldsymbol{(*)}$                    &                                                                                   \\ \hline
\end{tabular}

    \caption{Comparison among three placement procedures. Entries accompanied by a $\boldsymbol{(*)}$ represent contributions of this paper.  {Our surrogate approximation guarantee also extends seamlessly to the multi-SKU constrained-warehouse setting.}}
    \label{tab:intro_summary}
\end{table}

\linespread{\linespace}\selectfont{}

\subsection{Further Related Work \label{sec:literature}}

% In this section, we review a list of papers that are (closely or broadly) related to our work.

\textbf{Joint inventory placement and fulfillment.} \citet{chen2022approximation} studies the joint inventory placement and fulfillment problem with multiple items, obtaining a $(1-1/e)/4$ worst-case guarantee, and demonstrating its applicability at Anheuser Busch InBev (ABI). \citet{bai2022joint} study the joint inventory placement, promise, and fulfillment problem with multiple items and  warehouse capacity constraints, obtaining a $1/4$ worst-case guarantee. Their proposed inventory placement uses submodular optimization techniques to obtain an approximate solution for a fluid linear program. 
\citet{bai2022coordinated} studies the more general problem where the downstream decisions correspond to deciding an assortment, obtaining a $(1-1/e)/4$-approximation.
\citet{govindarajan2021joint} study the joint distribution and placement problem in the context of an omni-channel retailer. They provide an inventory placement heuristic that is, as the one we analyze, based on solving the optimal inventory placement problem assuming fulfillment will be hindsight-optimal, although they do not provide theoretical guarantees.    \citet{arlotto2023online} study the single-item joint inventory placement and fulfillment problem through the lens of regret minimization. They obtain sublinear regret for two state-of-the-art fulfillment policies, assuming that the initial inventory is optimal for the specific policy. {All of these models use the classic Temporal Independence demand model, whereas the results we obtain also hold in more general settings.} \citet{jasin2024inventory} study how Offline Placement performs in the cost minimization version of the joint placement and fulfillment problem. They provide several instances where Offline Placement can perform arbitrarily badly if coupled with Myopic Fulfillment, highlighting the importance of deploying this placement with a high-quality fulfillment policy.
% \bnote{Clarify that other placement procedures perform better. Explain why it doesn't contradict our findings.}\Willcomment{To be clear, this is referring to e.g. Fluid, not other placement procedures}, \Willedit{which does not contradict our findings because...}

\textbf{Other coordinated decisions in e-commerce retail.} The recent trend of studying coordinated decisions in e-commerce retail goes beyond inventory placement and fulfillment. \citet{lei2018joint} study the problem of jointly pricing and fulfillment for different items of a catalog during a fixed time horizon. \citet{lei2022joint} study the more general problem of joint product framing and order fulfillment under the multinomial logit model. They solve a deterministic approximation for the stochastic control problem and develop a randomized rounding scheme for this solution. \citet{jasin2022joint} studies a joint inventory replenishment, pricing, and fulfillment problem with multiple stores and warehouses. They develop a Lagrangian based heuristic that achieves sub-linear regret in the time horizon. \citet{devalve2023understanding} primarily studies the effects of flexibility in e-commerce retail supply chains, but their work takes a look at how several aspects interact. Through numerical experiments using real-world data, they study the interaction of fulfillment, placement, and network design. In particular, they find that if the deployed fulfillment policy is myopic, adding flexibility to the network can increase total costs. 
\cite{devalve2023approximate} study the two-stage problem of deciding a network structure followed by fulfillment decisions, which they assume can be made in an offline fashion, and derive constant factor guarantees. \citet{zhao2023managing} studies an inventory replenishment and fulfillment problem in a network structure that resembles the JD.com network that we study. 
{\cite{devalve2023cost} studies, among other problems, the multi-item joint inventory buying, placement, and fulfillment problem with the presence of fixed costs, where orders can contain several items. They show that it is NP-hard to approximate this problem up to a constant factor. {They also derive approximation guarantees for a broad range of problems.}}

\textbf{Newsvendor Networks.} \citet{van1998investment,mieghem2002newsvendor} study a multi-dimensional newsvendor model, closely related to optimizing inventory assuming that fulfillment will be hindsight optimal. The difference with inventory placement is that in the latter, the total number of units, $Q$, has already been decided exogenously. Closely related is \citet{govindarajan2021distribution}, where they study a multi-location newsvendor network model where the only information known about the joint demand distribution is the mean vector and covariance matrix. They take a distributionally robust approach to find the inventory decision that minimizes the worst-case expected cost of fulfilling demand. \citet{birge2025inventory} study the problem of purchasing and placing inventory units on a network before facing uncertain demand under the assumption that fulfillment will be hindsight optimal, deriving complexity and approximation results. For recent developments and an in-depth literature review related to Newsvendor networks, we refer the reader to \citet{devalve2023approximation}.

\section{Problem Statement\label{subsec:prob_statement}}
We study the two-stage problem faced by an e-commerce retailer (e-tailer) who has to make an inventory placement decision at the beginning of a time horizon (first stage), followed by sequential fulfillment decisions during that time horizon (second stage). The e-tailer manages a supply chain network consisting of warehouses $i\in [n]$ that can hold inventory and serve different demand nodes $j\in[m]$. Demand node $j$ can be thought of as a specific district or ZIP code that we call $j$. When demand from node $j$ arrives, the e-tailer must decide from which warehouse (if any) that demand is going to be satisfied. If the e-tailer decides to deliver the item from warehouse $i$, they collect a reward of $r_{ij}\in[0,1]$. (Here, we are assuming that the rewards are bounded and normalized to lie between 0 and 1.) {We say that warehouse $i$ can serve demand of node $j$ if $r_{ij}>0$. For each demand node $j\in[m]$, we define its degree $d_j := |\{i\in[n]: r_{ij}>0\}|$ as the number of warehouses that can serve the demand of this node. We further define $d:=\max_{j\in[m]} d_j$. Parameter $d$ is an indicator of fulfillment flexibility in the network, and our guarantees will depend on it.} {(Here we assume a single SKU, but we extend our results to the multi-SKU constrained-warehouse setting in \Cref{sec:multisku}.)}

\textbf{First stage problem -- Inventory placement.} In the inventory placement problem, the e-tailer must distribute an incoming shipment of $Q\in \bZ_{\geq0}$ inventory units across the $n$ distribution centers they manage, where $\bZ_{\geq0}$ are the non-negative integers. Formally, the e-tailer must decide on an inventory placement in $\X:= \{ x\in \bZ_{\geq0}^n : \sum_{i\in[n]} x_i = Q \}$.
% We say that the e-tailer chooses a placement procedure $P$: a function that maps the problem instances (to be specified below) to a (potentially randomized) inventory placement $x^P$.
{We note that $Q$ is not a decision variable since we assume that an upstream team has already decided the total quantity to stock.} {(Including the total quantity $Q$ to stock at a unit cost of $c$, and investigating the overall performance of the system under different surrogates is an interesting research direction.)} As a remark, all of our results directly extend to the setting where each warehouse has a capacity limiting the number of stock units that can be placed there, and where each warehouse also has a starting inventory. We omit these aspects for the sake of exposition, and refer the reader to \Cref{app:capacity} for a brief explanation on how to extend our algorithms to this setting.

\textbf{Second stage problem -- Fulfillment.} The fulfillment problem takes an inventory placement as input and faces a random stream of demand arriving from different nodes. We say that the total number of demand arrivals is $T$, which we call the time horizon and is generally random. We characterize the sequence of demand arrivals by a random vector $J = (j_t)_{t=1}^T$, where $j_t = j$ if and only if the $t$-th demand arrival comes from node $j\in [m]$. {This sequence is random, so it is unknown to the e-tailer, who only gets to observe demand arrivals one by one as they arrive.} Sequence $J$ follows a probability distribution called the \textit{demand model}. We make the standard assumption that demand is independent of our inventory placement and fulfillment decisions. We define $D_j = \sum_{t=1}^T \Ind\{j_t=j\}$ to be the total number of demand arrivals from node $j$ during the time horizon, and we further define $D = (D_j)_{j\in[m]}$ to be the vector containing these total demands. 
Our approach allows us to analyze and obtain results for a general family of random demand models. The only assumptions that we make are that the demand allows a fulfillment policy with a constant factor competitive ratio and that the total demand vector $D$ is exogenous to our placement and fulfillment decisions (this is further explained in \Cref{subsec:demand}).
The e-tailer has an initial inventory $x \in \X$ decided in the first stage, and starts observing the demand sequence one by one. When the $t$-th demand request arrives, the e-tailer observes its origin $j_t$, and must irrevocably decide at most one warehouse $i\in[n]$ with remaining inventory from which to fulfill the request. Once the warehouse is chosen, the e-tailer depletes one unit of inventory, collects a reward $r_{ij_t}$, and observes the next arrival.
{We say that these online decisions are prescribed by a fulfillment policy $\pi$, and let $\Pi$ denote the class of all admissible fulfillment policies that make decisions without knowing the future.} In its most general form, a policy $\pi$ decides from which warehouse to fulfill each demand arrival based on the node from where it arrives and the current inventory status together with the history of all previous demand arrivals and fulfillment decisions, and it is allowed to randomize. {Reductions to the policy class} can be made when further structure is imposed on the demand model.

An instance of the joint placement and fulfillment problem includes the total inventory to be placed $Q$, the sets of warehouses $[n]$ and demand nodes $[m]$, the rewards $(r_{ij})_{i,j\in[n]\times[m]}$, plus any demand parameters (concrete examples are provided in \Cref{subsec:demand}). We use $\ONL(x,\pi)$ to denote the expected reward collected by deploying an inventory placement $x$ and a fulfillment policy $\pi$, where the expectation is over the random demand and inner randomness of the inventory placement and fulfillment policy. We further define $\OPT(x):= \max_{\pi\in \Pi}  \ONL(x,\pi)$ as the expected reward obtained by the best possible policy for a given instance and initial inventory. The overall goal of the e-tailer is to maximize the expected total reward collected during the time horizon by jointly deciding on an inventory placement and a fulfillment policy. To decide on an inventory placement, the e-tailer uses a \textit{placement procedure} $P$ that takes in an instance and outputs a (potentially random) inventory placement $x^P\in\X$. To decide on a fulfillment policy, the e-tailer uses a \textit{fulfillment procedure} $F$ that takes in an instance along with an inventory placement {$x$} and outputs a fulfillment policy $\pi^F \in \Pi$. Our goal is to derive a placement procedure $P$ along with a fulfillment procedure $F$ such that
\[ \ONL(x^P,\pi^F)  \geq \gamma \max_{x\in\X} \sup_{\pi \in \Pi} \ONL(x,\pi) - \varepsilon =\gamma \max_{x\in\X} \OPT(x) - \varepsilon\]
for all instances, with \textit{approximation ratio} $\gamma\in[0,1]$ being as large as possible, and sampling error $\eps\ge0$ (to be specified) being as small as possible.

% since we will be working with the linear relaxation of $\max_{x\in \X} \OFF(x)$.

\section{Optimizing the Offline Surrogate \label{sec:techinical}}

A solution to the joint placement and fulfillment problem consists of both a placement and fulfillment procedure. This section focuses on deriving the placement procedure, which optimizes the Offline surrogate.

The Offline surrogate represents the expected reward from Offline fulfillment—a hypothetical (non-implementable) algorithm that knows demand $D$ in advance and makes optimal hindsight fulfillment decisions. Given $D$ and initial inventory $x$, the reward from Offline fulfillment is $\OFF(x,D)$, the optimal value of a bipartite matching LP defined in \Cref{subsec:theo_contributions}. The fulfillment decisions correspond to the optimal LP variables $y_{ij}$, representing the number of units delivered from warehouse $i$ to demand node $j$. Since this value depends only on $D$ (not the arrival order), $\OFF(x,D)$ serves as an upper bound on any online algorithm's reward, as online decisions also lie in the matching LP polytope. Slightly abusing notation, we define the Offline surrogate as $\OFF(x) := \E_D[\OFF(x,D)]$, averaging over demand realizations. Both $\OFF(x,D)$ and $\OFF(x)$ extend to fractional placements $x$ in the convex hull $\CH(\X) := \{ x\in\R^n_+: \sum_{i\in[n]} x_i = Q\}$, allowing us to consider relaxations where $x$ is fractional.

Ideally, we would solve $\max_{x\in\X} \OFF(x)$ exactly, but two challenges arise: integrality constraints in $\X$ and an exponentially large sample space for $D$. To address integrality, we use randomized rounding, solving the relaxed problem $\max_{x\in\CH(\X)} \OFF(x)$ and rounding to an integer while preserving a constant factor. To handle the large sample space, we apply sample-average approximation (SAA), sampling a small number of demand realizations and solving the SAA problem. The placement procedure consists of (1) sampling demand, (2) solving the SAA linear relaxation, and (3) randomly rounding the solution.

The rest of this section is organized as follows. In \Cref{subsec:rounding}, we describe how to round solutions of $\max_{x\in\CH(\X)}\OFF(x)$ while preserving a constant factor of the original value. In \Cref{subsec:statLearning}, we apply statistical learning theory to show that an arbitrarily good solution can be obtained from our sample-average approximation using a polynomial number of samples.

% The rest of this section is organized as follows. In \Cref{subsec:rounding} we explain how to round solutions of $\max_{x\in\CH(\X)}\OFF(x)$, obtaining at least a constant factor of the value of the original solution. In \Cref{subsec:statLearning} we use tools from statistical learning theory to show that we can obtain an arbitrarily good solution from our sample-average approximation using a polynomial number of samples.

\subsection{Randomized Rounding \label{subsec:rounding}}

A key step in our inventory placement procedure is to round a (potentially fractional) solution $x\in \CH(\X)$ without losing too much objective value when evaluating it in $\OFF$. To do so we apply the (star graph case of the) dependent rounding algorithm presented in \citet{gandhi2006dependent}. This algorithm takes a finite set $A$ and a weight $w_i\in[0,1] $ for every $i \in A$ as input. It outputs a vector of rounded random variables $W_i\in\{0,1\}$ for all $i\in A$ that satisfies three properties:
\begin{itemize}
    \item \textbf{Marginal Distribution}: $\E[W_{i}] = w_{i}$ for all $i\in A$.
    \item \textbf{Degree Preservation}:
    Let $\delta = \sum_{i\in A} w_{i}$. Then $\sum_{i\in A} W_i \in \{\lfloor \delta \rfloor, \lceil \delta \rceil\}$ with probability~1.
    \item \textbf{Negative Correlation}: For any subset $S\subseteq A$ and for any $b\in\{0,1\}$:
    \[ \PP\left(\bigcap_{i\in S} W_i = b\right) \leq \prod_{i\in S} \PP(W_i = b) .\]
\end{itemize}
We use this rounding algorithm as a subroutine, where we feed a solution $x\in\CH(\X)$ and obtain a rounded solution $R(x)\in \X$. In our application, we will only round the fractional layer of the solution $x\in\CH(\X)$, where we define the fractional part of $x_i$ as $x^f_i:=x_i - \flxi$. To round the solution, we input the set of warehouses, i.e.~$A = [n]$, with weights $w_i = x^f_i$ for all $i$ to the rounding algorithm by \citet{gandhi2006dependent}. Let $(W_i)_{i\in[n]}$ be the output of the rounding subroutine. We then set $R_i(x) = \flxi + W_i$ for all $i\in[n]$, where $R_i(x)$ is the $i$-th component of the rounded solution $R(x)$. In this application, the three properties of the randomized rounding algorithm by \citet{gandhi2006dependent} translate to:
\begin{itemize}[align=left, widest={\textbf{(P3)}},leftmargin=*]
    \item[\textbf{(P1)}] \textbf{Marginal Distribution}: $\E[R_i(x)]=x_i$  for all $i\in[n]$, \label{ppty:1}
    \item[\textbf{(P2)}] \textbf{Degree Preservation}: $\sum_{i\in [n]} R_i(x) = Q$ {and $R_i(x)\in \{\flxi, \flxi +1\}$ for all $i\in [n]$} with probability 1, \label{opty:2}
    \item[\textbf{(P3)}] \textbf{Negative Correlation}: For any subset $S\subseteq [n]$ and for any $b\in\{0,1\}$: \label{ppty:3}
    \[ \PP\left(\bigcap_{i\in S} R_i(x) = \flxi + b\right) \leq \prod_{i\in S} \PP(R_i(x) = \flxi + b) .\]
\end{itemize}

We establish the following guarantee for this rounding procedure on our problem.
\begin{theorem}\label{thm:rand_rounding} Consider an instance such that the maximum degree across demand nodes is $d$. For any fractional inventory placement $x\in CH(\X)$, and any rounding procedure $R$ that satisfies Properties \textnormal{(P1)-(P3)}, it holds that
\begin{align} \label{eqn:roundingGuarantee}
\E_R[ \OFF(R(x)) ] \geq \left(1 - \left(1-\frac{1}{d}\right)^d\right)\OFF(x).
\end{align}
\end{theorem}
To prove this, the first thing to notice is that $\bE_R[\OFF(R(x))]=\bE_{D,R}[\OFF(R(x),D)]$ ({because} $D$ is independent from $R$) and $\OFF(x)=\bE_D[\OFF(x,D)]$.  Therefore, to prove~\eqref{eqn:roundingGuarantee}, it suffices to show
\begin{align} \label{eqn:roundingGoal}
{\E_R[ \OFF(R(x),D^\samplePathOld) ] \geq \left(1 - \left(1-\frac{1}{d}\right)^d\right)\OFF(x, D^\samplePathOld)}
\end{align}
for any fixed realization of $D$. Given $D$, we will randomly construct a feasible solution to the LP defining $\OFF(R(x),D^\samplePathOld)$ that satisfies~\eqref{eqn:roundingGoal} in expectation over the randomly rounded inventory and the random construction of the solution. For the analysis, we will split the demand of each node $j$ into $D_j$ identical nodes with one unit of demand each. After doing so, we end up with the same problem except that now $D_j=1$ for all $j\in[T]$ under the new indexing.
(Note that we cannot split inventory units at warehouses in the same way, because that would change the parameter $d$ in our guarantee.) Hereafter, we fix a demand realization $D$, split $D$ into unit demand nodes, and let $(y_{ij}^\samplePathOld)_{(i,j)\in [n]\times [T]}$ refer to an optimal solution to the LP defining $\OFF(x,D^\samplePathOld)$ {(after the splitting of nodes so that $y_{ij}\in[0,1]$ for all $i,j$)}.

We use $(Z_{ij})_{(i,j)\in [n]\times[T]}$ to denote our proposed feasible solution for $\OFF(R(x),D)$. Its construction can be summarized in the following three steps.
\begin{enumerate}
    \item Round the feasible solution $x$ and obtain $R(x)$.
    
    \item For each warehouse $i\in [n]$, preliminarily assign its $R_i(x)$ inventory units across demand nodes $j\in[T]$. Let $Y_{ij}$ be an indicator for whether a unit from $i$ is preliminarily assigned to $j$. This step applies another iteration of the rounding algorithm from \citet{gandhi2006dependent}, with $Y_{ij}$ depending on whether $x_i$ was rounded up or down.
    %For each warehouse $i\in [n]$, determine a preliminary assignment of each of it's $R_i(x)$ inventory units across different demand nodes $j\in[T]$. We define $Y_{ij}$ to be the indicator that a unit of warehouse $i\in[n]$ was preliminary assigned to demand node $j\in[T]$. This step is performed with another iteration of the rounding algorithm by \citet{gandhi2006dependent}, and $Y_{ij}$ will depend on whether $x_i$ was rounded up or down.
    
    \item For each demand node $j\in[T]$, finalize assignments to construct $(Z_{ij})_{(i,j)\in [n]\times[T]}$. Demand node $j$ selects the preliminarily assigned unit yielding the highest reward, setting $Z_{ij} =1$ if and only if $i = \arg\max_{i:Y_{ij}=1} r_{ij}$, with arbitrary tie-breaking.
    %For each demand node $j\in[T]$, we determine a final assignment that we use as the feasible solution $(Z_{ij})_{(i,j)\in [n]\times[T]}$. To do this, demand node $j$ inspects all warehouses $i$ that had a unit preliminarily assigned to $j$, and chooses the inventory unit that provides the highest reward. We use $Z_{ij}$ to denote the indicator that $i$ was finally assigned to $j$, so $Z_{ij} =1 $ if and only if $i = \arg\max_{i:Y_{ij}=1} r_{ij}$ (and in case of ties we use an arbitrary tie-breaking rule).
\end{enumerate}
Step 1 is independent of $D$, while Steps 2 and 3 depend on $D$ and $R(x)$. In Step 3, each demand node selects at most one assigned unit, wasting the rest. This explains why the guarantee worsens as $d$ increases—larger $d$ means demand nodes can be served by more warehouses, increasing inventory waste.
% Step 1 is independent of $D$ by definition, whereas Steps 2 and 3 depend on $D$ and $R(x)$. Notice that in Step 3, at most one preliminary assigned inventory unit is selected by each demand node $j\in[T]$, and the remaining is wasted. Intuitively, this is why our guarantee decreases when $d$ increases. Large $d$ means that demand nodes can be served by many different warehouses, making it more likely to `waste' inventory units.

We now elaborate on Step 2: defining a preliminary assignment of the $R_i(x)$ inventory units of each warehouse $i$. Conditional on $R(x)$, this will be done independently for every $i\in[n]$, and to do this we again resort to the dependent rounding algorithm by \citet{gandhi2006dependent}. For each $i\in [n]$, we feed the rounding subroutine the set $[T]$ with weights that depend on whether $W_i=1$ or $W_i=0$. We denote these weights by $(y_{ij}^\HH)_{j\in [T]}$ (for high inventory, $W_i=1$) and $(y_{ij}^\LL)_{j\in [T]}$ (for low inventory, $W_i=0$). For the analysis to hold, we require that for all $i\in[n]$, these weights (probabilities) satisfy the following constraints:
\begin{align}
    y_{ij}^{\samplePathOld\HH}x_i^f + y_{ij}^{\samplePathOld\LL}(1-x_i^f)& = y_{ij}^\samplePathOld&\quad \forall j\in [T],\label{eq:marginal_dist_y}\\
    0\leq y_{ij}^{\samplePathOld\LL} \leq y_{ij}^{\samplePathOld\HH} & \leq 1& \quad \forall j\in [T],\label{eq:neg_corr_y}\\
    \sum_{j\in[T]} y_{ij}^{\samplePathOld\LL}  &\leq \flxi,\label{eq:inv_constraint1}\\
    \sum_{j\in[T]} y_{ij}^{\samplePathOld\HH}  &\leq \flxi+1.\label{eq:inv_constraint2}
\end{align}
\Cref{eq:marginal_dist_y} will make $\E[Y_{ij}^{\samplePathOld}] = y_{ij}^{\samplePathOld}$ for all $j\in [T]$. \Cref{eq:neg_corr_y} {is a logical monotonicity constraint that allows to transfer the negative correlation property of the rounding $(W_i)_{i\in[n]}$ to the preliminary assignment $(Y_{ij})_{i\in[n]}$, for every $j\in[T]$.}
\Cref{eq:inv_constraint1,eq:inv_constraint2} will make it so that we can always apply the procedure of \citet{gandhi2006dependent} without assigning more inventory units than what we have in hand, regardless of whether $R_i(x)=\lfloor x_i\rfloor$ or $R_i(x)=\lfloor x_i\rfloor+1$. We note that naive ways of setting these probabilities such as scaling, sampling, or splitting fail:
\begin{enumerate}
    \item \textbf{Scaling}: {Setting $y^\HH_{ij}=y_{ij}(\lceil x_i\rceil/x_i),y^\LL_{ij}=y_{ij}(\lfloor x_i\rfloor/x_i)$ for each $j$ satisfies all the conditions except that it could create values of $y^\HH_{ij}$ that are greater than 1 and hence invalid probabilities.}
    \item \textbf{Proportional Sampling}: {Another plausible strategy is to sample $R_i(x)$ times without replacement from $[T]$, with probabilities proportional to $y_{i1},\ldots,y_{iT}$.  However, this does not preserve marginals---take for example $y_{i1}=0.9,y_{i2}=0.6,x_i=1.5$.  This procedure would induce $y^\HH_{i1}=1$ and $y^\LL_{i1}=3/5$, but does not satisfy $0.5y^\HH_{i1}+0.5y^\LL_{i1}=y_{i1}$.}
    \item \textbf{Further Splitting}: {Naive methods would work if we could split warehouses so that $x_i\le1$ for all $i$; however, this would change the value of $d$ that we are parametrizing by.}
\end{enumerate}

Given the failure of these naive methods, we instead show fundamentally that a feasible adjustment $(y^\LL_{ij},y^\HH_{ij})_{j\in[T]}$ exists.
%which essentially boils down to applying the simple fact that $\lceil x_i\rceil - x_i\le\sum_j(\lceil y_{ij}\rceil - y_{ij})$ when $\sum_j y_{ij}=x_i$.
\begin{lemma}\label{lem:weightsExist}
    For all $i\in [n]$, there exist vectors of probabilities $(y_{ij}^{\samplePathOld\HH})_{j\in[T]}$ and $(y_{ij}^{\samplePathOld\LL})_{j\in[T]}$ that satisfy \Cref{eq:marginal_dist_y,eq:neg_corr_y,eq:inv_constraint1,eq:inv_constraint2}, {assuming that $\sum_{j\in[T]} y_{ij}\le x_i$ and $y_{ij}\in[0,1]$ for each $j\in[T]$}.
\end{lemma}

The proof of this lemma is deferred to \Cref{app:proof_weights_lemma}.

We proceed by lower-bounding the reward that can be obtained from each demand node by our proposed solution $(Z_{ij})_{(i,j)\in[n]\times[T]}$, in expectation over the randomized placement and assignment, in a way that allows us to relate it to the objective value of $\OFF(x,D)$.

\begin{lemma} \label{lem:ZBound}
Fix a realization of the demand vector $D$.  It holds for all demand nodes $j\in[T]$ that
\[ \E\left[\sum_{i\in[n]} r_{ij}Z_{ij}^\samplePathOld \right] \geq \left(1 - \left(1 - \frac{1}{d_j} \right)^{d_j}  \right)\sum_{i\in[n]} r_{ij} y_{ij}^\samplePathOld, \]
where the expectation is taken over the randomized rounding $R(x)$ and random assignments determined by $Y_{ij}$ and $Z_{ij}^\samplePathOld$.
\end{lemma}

We prove \Cref{lem:ZBound} in \Cref{app:proof_Zbound}, noting that a similar proof has appeared in \citet{brubach2021improved}. With these results, we can complete the proof of \Cref{thm:rand_rounding}, presented in \Cref{app:proof_rounding}. Importantly, we note that the approximation from \Cref{thm:rand_rounding} is tight.

\begin{lemma}\label{lem:tight_example}
    There exists a family of instances for which
    \[ \frac{\max_{x\in \X} \OFF(x)}{\max_{x\in \CH(\X)} \OFF(x)} \leq 1 - \left( 1 - \frac{1}{d}\right)^d. \]
\end{lemma}
The proof of this lemma is deferred to \Cref{app:tight_example}.

We close this subsection by noting another approach to approximating $\max_{x\in\X} \OFF(x)$: submodular optimization. However, this fails to yield better approximations for small $d$.
\citet{bai2022joint} show that $\OFF(x,D)$ is submodular over $x\in\X$ for fixed $D$. Since submodularity is preserved under convex combinations, $\OFF(x)$ is also submodular over $x\in\X$. Moreover, $\OFF(x)$ is non-decreasing in $x$, allowing a $(1-1/e)$-approximation of $\max_{x\in\X}\OFF(x)$ via greedy inventory placement \citep{nemhauser1978analysis}. More precisely, the submodular approach gives an approximation of $(1-(1-1/Q)^Q)$, where $Q$ is the inventory units placed. In applications, $Q$ is large, making this close to $1-1/e$. This bound cannot improve even if $d$ is small, as shown in \Cref{sec:greedySucks}. Another advantage of rounding over submodular optimization is that the $1-(1-1/d)^d$ randomized rounding guarantee extends to a general multi-SKU setting (\Cref{sec:multisku}), whereas submodular techniques only ensure $1/2$.
% We close this subsection by pointing out another possible approach for approximating $\max_{x\in\X} \OFF(x)$ which is to use submodular optimization. This approach, however, fails to get an approximation that improves for small $d$.
% \citet{bai2022joint} show that $\OFF(x,D)$ is a submodular function over $x\in\X$ for any fixed $D$. Since submodularity is preserved under a convex combination, $\OFF(x)$ is also a submodular function over $x\in\X$. It is straightforward to verify that $\OFF(x)$ is also a non-decreasing function of $x$, implying that we could obtain a $(1-1/e)$-approximation of $\max_{x\in\X}\OFF(x)$ by constructing an inventory placement greedily \citep{calinescu2007maximizing}. More precisely, the submodular approach would yield an approximation of $(1-(1-1/Q)^Q)$, where $Q$ is the number of inventory units to be placed. In applications $Q$ is large, so this approximation would be close to $1-1/e$. This factor cannot be further improved even if $d$ is small, as shown in \Cref{sec:greedySucks}. Another advantage of using rounding instead of submodular optimization techniques is that the $1-(1-1/d)^d$ randomized rounding guarantee holds in a more general, multi-SKU setting, as we show in \Cref{sec:multisku}. Submodular optimization techniques, on the other hand, would only lead to a guarante of $1/2$.

\subsection{Sample Average Approximation \label{subsec:statLearning}}

The rounding results in \Cref{subsec:rounding} are contingent on being able to solve
the problem $\max_{x\in\CH(\X)}  \OFF(x)$.
However, since $\OFF(x)$ takes an expectation over $D$ and the number of possible realizations of $D$ grows exponentially under a general demand model, the problem cannot be solved exactly.
Instead, we solve a sample-average approximation of the problem $\max_{x\in\CH(\X)}  \OFF(x)$, and prove that the sample-average approximation is not too sensitive to the samples drawn.

To elaborate, we independently sample $K$ vectors $D^1,\dots,D^K$ for the total demand according to the given demand model.  We define
\[ \hOFF(x) = \frac{1}{K} \sum_{k=1}^K \OFF(x,D^k), \]
and solve for $\hat{x}$, an optimal solution to the problem $\max_{x\in\CH(\X)}  \hOFF(x)$. In general, we will use the symbol $\wedge$ to refer to anything that depends on the samples $D^1,\dots,D^K$, and we will use $\E_\wedge$ when we are taking an expectation over the random samples. It will also be understood that the sample consists of exactly $K$ demand realizations. 

We can solve $\max_{x\in\CH(\X)}  \hOFF(x)$ by solving the following linear program:
\begin{align*}
    \max_{x\in \R^n_+, y\in\R^{[n]\times[m]\times[K]}_+} \quad & \frac{1}{K}\sum_{k=1}^K\sum_{i\in[n]}\sum_{j\in[m]} r_{ij}y_{ij}^k\\
    \mathrm{s.t.}\quad &\sum_{i=1}^n y_{ij}^k \leq D_j^k \quad &\forall j\in[m],k\in[K],\\
     &\sum_{i=j}^m y_{ij}^k \leq x_i \quad &\forall i\in[n],k\in[K],\\
     & \sum_{i=1}^n x_i = Q.\\
\end{align*}
Notice that this problem has a polynomial number of variables and constraints if $K$ is polynomial in the instance parameters. By contrast, the problem $\max_{x\in\CH(\X)}  \OFF(x)$ would have required an exponentially-sized $K$ to capture the support of $D$.
% Thus, the sampling step is crucial so that we can efficiently solve the optimization problem.

The main result of this subsection, which will allow us to use the sample-average solution $\hat{x}$ to obtain our approximation result, is the following.
\begin{theorem}\label{thm:stat_learning} For a random sample of $K$ IID demand realizations, we have
\[ \E_\wedge\left[\sup_{x\in \CH(\X)}   \hOFF(x) - \OFF(x) \right] = O\left( Q\sqrt{\frac{ n \log n}{K}}\right). \]
\end{theorem}
{\Cref{thm:stat_learning} provides a uniform convergence guarantee on the generalization error of how much $\hOFF$ can overestimate the value of any solution $x$ and its proof uses tools from statistical learning theory.
We note that the space of fractional solutions $\CH(\X)$ is continuous and infinite; however, we can apply a vector contraction inequality after showing the function $\OFF(x)$ to be Lipschitz in $x$.}{ The proof of \Cref{thm:stat_learning} is presented in \Cref{app:proof_stat_learn_thm}.}

\section{Approximating the Joint Problem \label{sec:joinproblem}}

In this section, we build on the results from \Cref{sec:techinical} to derive
% \Willdelete{our main theoretical result:}
a placement procedure and fulfillment procedure that jointly achieve an $\alpha (1-(1-1/d)^d)$-approximation of the optimal joint solution. {This can only be possible if the demand model admits a fulfillment approximation guarantee in the first place, as we describe in \Cref{subsec:demand}.}
% In \Cref{subsec:demand}, we list and discuss the assumptions we make about the demand model and provide a list of examples.
In \Cref{subsec:mainResult}, we present the joint result, provide a sketch of its proof, and provide some remarks and discussion.

% In this section, we present and develop our main theoretical result:  For the fulfillment procedure, we output the $\alpha$-competitive fulfillment policy that we assume to exist. Therefore, this section is mostly about the placement procedure, which is based on optimizing the Offline surrogate. Ideally, we would aim to solve $\max_{x\in\X} \OFF(x)$ exactly. However, we encounter with two main challenges when attempting this: the integrality constraints in $\X$, and the exponentially sized sample space for random variable $D$. To overcome the first challenge we take a randomized rounding approach, where we instead solve the linear relaxation $\max_{x\in\CH(\X)} \OFF(x)$ and randomly round the optimal solution to make it integer without losing more than a constant factor. To overcome the second challenge we take a sample-average approximation approach, where we sample a small number of demand realizations and work with the SAA problem instead. The placement procedure can then be summarized as 1) sampling demand realizations, 2) solving the SAA version of the linear relaxation of the Offline surrogate, and 3) randomly rounding the solution obtained in step 2).

\subsection{Assumptions on Demand Model \label{subsec:demand}}

As mentioned in \Cref{subsec:prob_statement}, our analysis applies to a broad class of demand models, defined as a distribution over sequences of demand nodes (of random length). We assume the following.
\begin{assumption}\label{assumption:demand}
The demand model is such that:
\begin{itemize}
    \item[(a)] for any instance, and any starting inventory $x\in \X$, we can efficiently compute a fulfillment policy $\pi^\alpha$ such that $\ONL(x,\pi^\alpha) \geq \alpha \OFF(x)$, \label{assumpt:competitive}
    \item[(b)] the total demand vector $D$ {does not depend on} the placement and fulfillment policy, and 
    \label{assumpt:indep}

    \item[(c)] {a sample of the total demand vector $D$ can be drawn in polynomial time.}
    % the total demand vector $D$ can be sampled efficiently.
    \label{assumpt:sampling}
\end{itemize}
\end{assumption}
% The first assumption is necessary to carry out our analysis. (It can technically be relaxed to be satisfied only for inventory placements that can be the output of our proposed placement procedure.) The second assumption is standard. We remark that the independence from our decisions only has to hold for the total demand, and the order in which demand arrives can be endogenous/adversarial as long as the remaining assumptions hold. The third assumption is required because our proposed placement procedure uses samples of $D$ as a subroutine.
Assumption (a) is necessary for our analysis (though it could be relaxed for placements generated by our procedure). Assumption (b) is standard; independence applies only to total demand, allowing adversarial arrival orders. Assumption (c) is required as our placement procedure relies on sampling $D$.

We mention two demand models in the literature that satisfy these assumptions as examples.

\textbf{Temporal Independence model.} A classic model in revenue management and online matching \citep{alaei2012online}, where the time horizon $T$ is deterministic and known, and each demand node $j$ arrives independently at time $t$ with probability $p_{tj}$. Some time steps may have no arrivals, represented by a dummy demand node with zero reward. Here, $\sum_{j\in[m]} p_{tj}=1$ for all $t\in[T]$. Assumptions (a), (b), and (c) hold, with $\alpha = 1/2$ \citep{alaei2012online}, as $D$ follows a multinomial distribution with $T$ trials.

% \textbf{Temporal Independence model.} This is a classical model in {revenue management and online matching} (e.g., \citet{alaei2012online}). Here, the time horizon $T$ is deterministic and known to the e-tailer. In each time step we will have $j_t = j$ with probability $p_{tj}$, where $j_t$ is independent of $j_{t'}$ if $t\neq t'$. The model allows the occurrence that no demand arrives in a given time step. This is represented by a dummy demand node that has zero reward and hence is always rejected. Then, for all $t\in[T]$ we have $\sum_{j\in[m]} p_{tj}=1$. The demand parameters in this model are the time horizon $T$ and arrival probabilities $(p_{tj})_{t,j\in[T]\times[m]}$. Assumption (a) is satisfied with $\alpha = 1/2$ \citep{alaei2012online}. Assumption (b) is satisfied by definition, and Assumption (c) is satisfied because $D$ corresponds to a multinomial variable with $T$ trials and $m$ mutually exclusive events.

\textbf{Spatial Independence model.} Recently introduced by \citet{aouad2022nonparametric}, this model samples each node’s total demand $D_j$ from independent distributions $(G_j)_{j\in[m]}$. An adversary then determines the arrival order to minimize the e-tailer's reward. Here, $T = \sum_{j=1}^m D_j$ is random and unknown. Assumptions (a), (b), and (c) hold, with $\alpha = 1/2$ \citep{aouad2022nonparametric}, provided each $G_j$ can be sampled efficiently.

% \textbf{Spatial Independence model.} In this model, recently introduce by \citet{aouad2022nonparametric}, the total demand $D_j$ of each node $j\in[m]$ is sampled from independent distributions $(G_j)_{j\in[m]}$. Once the total demands are sampled and the e-tailer has chosen its policy, an adversary is allowed to choose the order in which the requests will arrive such that the reward collected by the e-tailer is minimized. Note that the time horizon $T=\sum_{j=1}^m D_j$ is now random and unknown to the e-tailer. Assumption (a) is again satisfied with $\alpha = 1/2$ \citep{aouad2022nonparametric}. Assumption (b) is true by definition, and Assumption (c) holds as long as each one of $(G_j)_{j\in[m]}$ can be sampled efficiently.

 \subsection{Approximation for the Join Fulfillment and Placement Problem \label{subsec:mainResult}}

% We proceed by presenting the joint placement and fulfillment solution. The placement procedure we deploy is $R(\hat{x})$, which was developed in \Cref{sec:techinical} and is computed following these three steps:
% \begin{enumerate}
%     \item Sample $K$ IID demand realizations $D_1,\dots,D^K$.
%     \item Solve for $\hat{x}$, an optimal solution of $\max_{x\in\CH(\X) }\widehat{\OFF}(x)$.
%     \item Randomly round $\hat{x}$ using $R$, obtaining $R(\hat{x})$.
% \end{enumerate}
% The placement procedure we use is the one that outputs $\pi^\alpha$, which exists by \Cref{assumpt:competitive} and is $\alpha$-competitive.

We now present the joint placement and fulfillment solution. The placement procedure, $R(\hat{x})$, follows three steps: \begin{enumerate} \item Sample $K$ IID demand realizations $D_1,\dots,D_K$. \item Solve for $\hat{x}$, the optimal solution of $\max_{x\in\CH(\X)}\widehat{\OFF}(x)$. \item Randomly round $\hat{x}$ using $R$ to obtain $R(\hat{x})$. \end{enumerate} The fulfillment policy used is $\pi^\alpha$, which exists by \Cref{assumpt:competitive} and is $\alpha$-competitive.
We obtain the following guarantee:
\begin{theorem}\label{thm:main}
         For any instance where at most $d$ warehouses serve a single demand node, we efficiently obtain an inventory placement $R(\hat{x})$ and a fulfillment policy $\pi^\alpha$ satisfying
    \[\E_{\wedge,R}\left[\ONL(R(\hat{x}),\pi^\alpha) \right] \geq \alpha\left(1 - \left(1 - \frac{1}{d}\right)^d\right)\max_{x\in\X}\OFF(x) - \varepsilon, \]
    using $K = O\left(\frac{Q^2n\log n}{\varepsilon^2}\right)$ demand samples.
    \end{theorem}
% Since $\OFF(x)\geq \OPT(x)$ for all $x\in \X$, \Cref{thm:main} implies that this joint solution provides an $\alpha(1-(1-1/d)^d)$-approximation for the joint placement and fulfillment problem. The proof of this theorem is deferred to \Cref{app:proof_main}, and is inspired by the three-step approximation framework by \citet{bai2022joint,bai2022coordinated}. 
Since $\OFF(x)\geq \OPT(x)$ for all $x\in \X$, this provides an $\alpha(1-(1-1/d)^d)$-approximation for the joint problem. The proof (deferred to \Cref{app:proof_main}) builds on the three-step approximation framework of \citet{bai2022joint,bai2022coordinated}, which involves: 1) defining a surrogate function $f$ such that $f(x) \geq \OPT(x)$ for all $x\in\X$, 2) finding an inventory placement $x'$ with $f(x')\geq \beta \max_{x\in\X} f(x)$, and 3) designing a fulfillment policy $\pi$ such that $\ONL(\pi,x') \geq \alpha f(x')$.This yields an $\alpha\beta$ approximation: 

% This framework consists of 1) coming up with a surrogate function $f$ such that $f(x) \geq \OPT(x)$ for all $x$, 2) obtaining an inventory placement $x'$ such that $f(x')\geq \beta \max_{x\in\X} f(x)$, and 3) coming up with a fulfillment policy $\pi$ such that $\ONL(\pi,x') \geq \alpha f(x')$. This way, they can obtain an $\alpha\beta$ approximation by using $x'$ and $\pi$:
\begin{align}
    \ONL(\pi,x') \geq \alpha f(x') \geq  \alpha \beta \max_{x\in\X} f(x)  \geq   \alpha  \beta\max_{x\in\X} \OPT(x). \label{eq:3step}
\end{align}
In our case, the surrogate function is $\OFF$. For step 2, we apply randomized rounding, achieving the $(1-(1-1/d)^d)$-approximation from \Cref{thm:rand_rounding}. To handle the large demand space, we use sample-average approximation. Since this prevents a direct application of \Cref{eq:3step}, we incorporate learning theory techniques. Finally, step 3 follows from the assumed existence of the $\alpha$-competitive fulfillment policy.
% In our proof, the surrogate function of step 1) is $\OFF$. For step 2), we use randomized rounding and obtain the $(1-(1-1/d)^d)$-approximation from \Cref{thm:rand_rounding}. To overcome the exponentially sized support of the demand vector $D$ we resort to sample-average approximation. This extra sampling does not allow us to use the clean chain of inequalities in \Cref{eq:3step}, so {here we deviate from the three-step approximation framework and apply some learning theory.}

We would like to remark on the advantage of using $\OFF(x)$ as the surrogate function instead of using a fluid approximation, approach used by \citet{bai2022joint}. Let $\FLU(x)$ be the value of this fluid approximation with inventory placement $x$, defined as
\begin{align*}
\FLU(x):= \max_{y\geq0} \quad & \sum_{i\in[n]}\sum_{j\in[m]} y_{ij} r_{ij}\nonumber\\
\mathrm{s.t.}\quad & \sum_{i\in [n]} y_{ij} \leq \E[D_j] \quad \forall j\in[m], \\
&\sum_{j\in[m]} y_{ij}\leq x_i \quad \forall i\in[n].
\end{align*}
That is, $\FLU(x)$ is equal to $\OFF(x,\E[D])$. Since $\OFF(x,D)$ is concave in $D$, Jensen's inequality implies that $\FLU(x) \geq \OFF(x)$ for all $x\in \CH(\X)$. Hence, to carry out our same analysis we would require that $\pi^\alpha$ satisfies the stronger assumption
\begin{equation}
    \ONL(x,\pi^\alpha) \geq \alpha \FLU(x) \quad \forall x\in \X. \label{eq:fluaprox}
\end{equation}
In fact, if we consider demand models that additionally satisfy this assumption, we obtain a strictly smaller subset than we would get only imposing \Cref{assumpt:competitive}. For instance, the Spatial Independence model described in \Cref{subsec:demand} does not allow any policy that satisfies \Cref{eq:fluaprox} with $\alpha$ bounded away from 0, but it does admit a policy that satisfies \Cref{assumpt:competitive} with $\alpha = 1/2$ \citep{aouad2022nonparametric}. Thus, our approach does not only obtain a strictly better competitive ratio, but it also works for a strictly larger set of demand models.

We highlight the practical aspects of our placement procedure in \Cref{thm:main}. First, it is entirely sample-driven, allowing past demand data to be used without fitting an exact distribution. Second, it requires only aggregate demand samples over time, independent of arrival order. Third, it is easily obtained by solving a linear program, with rounding needed only for rare fractional solutions. When the optimal solution is integer, our guarantee improves to $\alpha$.

% We would also like to highlight the practical aspects of our placement procedure in \Cref{thm:main}. First, it is completely sample-driven. In practice, past demand can be used as demand samples, and thus there is no need to fit an exact demand distribution. Second, our procedure only required samples of aggregate demand over the time horizon, so it is not reliant on knowing the order in which each customer from each location arrived. Third, it is straightforward to obtain by solving a linear program and potentially needing to round the solution if it turns out to be fractional. We have found that fractional solutions are a rare occurrence, and whenever the optimal solution of the linear program is integer, our guarantee improves to $\alpha$.

All in all, \Cref{thm:main} formalizes that offline fulfillment yields strong outcomes when paired with an effective fulfillment policy. Our placement procedure optimizes as if the fulfillment team knew the future exactly. The result shows this approach achieves state-of-the-art approximation guarantees when fulfillment is highly competitive relative to offline. This aligns with intuition—a strong fulfillment team allows the placement team to be optimistic about effective inventory distribution.

% All in all, \Cref{thm:main} formalizes the notion that assuming offline fulfillment has a good outcome if a good fulfillment policy is being deployed. {To elaborate, our placement procedure, which solves the offline relaxation to determine the inventory placement, is optimizing as if the downstream fulfillment team can make decisions as if they knew the future exactly.  Our result shows that this leads to state-of-the-art approximation guarantees when the fulfillment procedure is indeed quite good, in the sense of having a competitive ratio relative to offline.  This intuition makes sense---a good fulfillment team means that the placement team can be optimistic when making assumptions about how well the placed inventory will be distributed.

\section{Extension of Randomized Rounding to Multiple SKUs Setting \label{sec:multisku}}

We now change our focus to
%study the problem of optimizing inventory placement using the Offline surrogate in 
the multi-SKU setting of \citet{bai2022joint}. We show how to apply our result for the single-SKU setting in this case without loss in performance, and therefore obtain state-of-the-art guarantees for approximating the inventory placement problem under the Offline surrogate in the presence of multiple SKUs.

\subsection{Problem Setting \label{subsec:multiSetting}}

In the multi-SKU setting, we have $s$ different SKUs, indexed by $\ell \in [s]$. Each warehouse $i$ is equipped with an initial inventory of $x_{i\ell}$ units of SKU $\ell$. To avoid introducing additional notation, we will again use $x$ to denote this (now) $n\times s$ inventory matrix, and use $x^\ell:=(x_{i\ell})_{i \in [n]}$ to denote its $\ell$-th column (associated to SKU $\ell$). Random demand is now represented by an $m\times s$ matrix $D$, where $D_{j\ell}$ are the units of SKU $\ell$ demand originating from demand node $j$. Let $D^\ell:=(D_{j\ell})_{j\in[m]}$ represent the $\ell$-th column of the demand matrix. There is a reward of $r_{ij}^\ell$ for fulfilling a unit of demand for SKU $\ell$ originating from demand node $j$ using inventory in warehouse $i$. (In this problem, we do not account for potential reward gains from package consolidation. This is consistent with \citet{bai2022joint}.) Given an initial inventory matrix $x$ and random demand realization matrix $D$, the value of offline fulfillment is given by the following linear program
\begin{align}
\OFFM(x,D)= \max_{y\geq0} \quad & \sum_{\ell\in[s]}\sum_{i\in[n]}\sum_{j\in[m]} y_{ij}^\ell r_{ij}^\ell\nonumber\\
\mathrm{s.t.}\quad & \sum_{i\in [n]} y_{ij}^\ell \leq D_{j\ell} \quad \forall j\in[m],\,\forall \ell \in [s], \label{const:demandSideMulti}\\
&\sum_{j\in[m]} y_{ij}^\ell\leq x_{i\ell} \quad \forall i\in[n],\,\forall \ell \in [s].\label{const:supplySideMulti}
\end{align}
Here, Constraint (\ref{const:demandSideMulti}) says that for each SKU and each demand node, we cannot fulfill more than the demand originating from said node. Constraint (\ref{const:supplySideMulti}) says that we cannot use more than the available inventory for any warehouse and SKU pair.  As in the single-SKU setting, the value of the Offline surrogate corresponds to the expected value of the Offline fulfillment LP, taken over the random demand distribution: $\OFFM(x) = \E_D[\OFFM(x,D)]$.
{Note that the optimization problem defining $\OFFM(x,D)$ can be decoupled across SKUs $\ell$.}

We are interested in finding an optimal inventory placement for the Offline surrogate in the multi-SKU setting. Here, for each SKU $\ell$, we let $Q_\ell$ denote the number of inventory units to be placed. Each warehouse $i$ has a capacity $C_i$ that is crucially shared across SKUs, {coupling the inventory placement decisions across $\ell$}. Again, extending the notation from previous sections, we define the set of feasible inventory placements as
\begin{align*}
    \XM = \{ x\in \N^{n \times s}: &\sum_{i\in [n]} x_{i\ell} = Q_\ell\quad \forall \ell\in[s],\\
    &\sum_{\ell \in [s]} x_{i\ell} \leq C_i \quad \forall i\in [n]\}.
\end{align*}
We also define the convex hull of the feasible placements set $\CH(\XM)$ as $\XM$ after relaxing the integrality constraints. Finding an optimal inventory placement for the Offline surrogate formally corresponds to solving $\max_{x\in \XM} \OFFM(x)$. This is a hard combinatorial problem, and therefore we resort to obtaining good approximations.

\subsection{Rounding a Fraction Solution \label{subsec:roundingMulti}}

We proceed by showing how to extend our randomized rounding analysis from \Cref{subsec:rounding} to obtain state-of-the-art approximation guarantees for the Offline surrogate {in the multi-SKU setting}.

As we did in the single-SKU setting, we solve the linear relaxation $\max_{x\in \CH(\XM)} \OFFM(x)$ and round the optimal solution obtained.  To round the solution, we now apply the general bipartite graph case of the dependent rounding algorithm presented in \citet{gandhi2006dependent}. This algorithm takes a bipartite graph $G=(A,B,E)$ and weights $w_{i\ell}\in[0,1] $ for every edge $(i,\ell) \in E$ as input. It outputs a vector of random variables $W_{i\ell}\in\{0,1\}$ for all $(i,\ell)\in E$ that satisfies three properties:
\begin{itemize}
    \item \textbf{Marginal Distribution}: $\E[W_{i\ell}] = w_{i\ell}$ for every edge $(i,\ell)\in E$.
    \item \textbf{Degree Preservation}:
    For any node $i\in A \cup B$ define $\Delta(i)$ the set of {nodes adjacent to it, and let $\delta_i = \sum_{\ell \in \Delta(i)} w_{i\ell}$ be its fractional degree. Then, for any node $i\in A \cup B$ we have $\sum_{\ell \in \Delta(i)} W_{i\ell} \in \{\lfloor \delta_i \rfloor, \lceil \delta_i \rceil\}$ with probability~1.}
    \item \textbf{Negative Correlation}: For any node $i\in A \cup B$, any subset $S\subseteq \Delta(i)$ of {nodes} adjacent to $i$, and for any $b\in\{0,1\}$:
    \[ \PP\left(\bigcap_{\ell\in S} W_{i\ell} = b\right) \leq \prod_{\ell\in S} \PP(W_{i\ell} = b) .\]
\end{itemize}
We use this rounding algorithm as a subroutine, where we feed a solution $x\in\CH(\XM)$ and obtain a rounded solution $R(x)\in \XM$. In our application, we will only round the fractional layer of the solution $x\in\CH(\XM)$, where we define the fractional part of $x_{i\ell}$ as $x^f_{i\ell}:=x_{i\ell} - \flxil$.
To round the solution, we give the following input to the rounding algorithm by \citet{gandhi2006dependent}. The bipartite graph will have the set of warehouses on one side and the set of SKUs on the other side, and it will be complete. That is, $G = ([n], [s], [n]\times [s])$. For the weights, we use the fractional layer of our linear relaxation: $w_{i\ell} = x^f_{i\ell}$ for all $(i,\ell)\in [n]\times[s]$. Let $(W_{i\ell})_{(i,\ell)\in[n]\times [s]}$ be the output of the rounding subroutine. We then set $R_{i\ell}(x) = \flxil + W_{i\ell}$ for all $(i,\ell)\in[n]\times [s]$, where $R_{i\ell}(x)$ is the $(i,\ell)$-th entry of the rounded solution matrix $R(x)$. In this application, the marginal distribution property translates to $\E[R_{i\ell}(x)] = x_{i\ell}$ for all $(i,\ell)\in[n]\times [s]$. The degree preservation property implies that $\sum_{i\in [n]} R_{i\ell}(x) = Q_\ell$ and  $\sum_{\ell\in[s]} R_{i\ell}(x) \leq C_i$ with probability 1.

The main result of this section is that the rounding guarantee of \Cref{thm:rand_rounding} holds in the multi-SKU setting.
\begin{corollary}\label{cor:rand_roundingMulti} For any instance with multiple SKUs such that the maximum degree across demand nodes is $d$ and for any fractional inventory placement $x\in CH(\XM)$, it holds that
\begin{align} \label{eqn:roundingGuaranteeMulti}
\E_R[ \OFFM(R(x)) ] \geq \left(1 - \left(1-\frac{1}{d}\right)^d\right)\OFFM(x).
\end{align}
\end{corollary}

The key observation to prove this corollary is that $\OFFM(x)$ can be decomposed into $s$ subproblems, one for each SKU. Indeed, we have
\begin{align*}
    \OFFM(x)=E_D[\OFFM(x,D)] = \sum_{\ell \in [s]} \E_D[\OFF(x^\ell, D^\ell)] = \sum_{\ell \in [s]} \E_{D^\ell}[\OFF(x^\ell, D^\ell)] = \sum_{\ell \in [s]} \OFF_\ell(x^\ell), 
\end{align*}
where we implicitly define $\OFF_\ell(x^\ell):= E_{D^\ell}[\OFF(x^\ell,D^\ell)]$. Therefore we have
\[ \E_R[ \OFFM(R(x)) ] =  \sum_{\ell \in [s]} \E_R[\OFF_\ell(R^\ell(x))],\]
where $R^\ell(x)$ is the $\ell$-th column of the rounded inventory matrix. In order to obtain the corollary, it suffices to show that
\begin{equation}
    \E_R[\OFF_\ell(R^\ell(x))] \geq \left(1-\left(1-1/d\right)\right)\OFF_\ell(x^\ell) \label{eq:sufficientCondMulti}
\end{equation}
 for any $\ell\in[s]$. In order to do that, we simply notice that $R^\ell(x)$ satisfies Properties (P1)-(P3), the hypotheses of \Cref{thm:rand_rounding}. Therefore, we can invoke this theorem and establish (\ref{eq:sufficientCondMulti}), which ultimately implies \Cref{cor:rand_roundingMulti}.

The approximation guarantee in \Cref{cor:rand_roundingMulti} is at least $1-1/e$, which improves upon the state-of-the-art guarantee of $1/2$ obtained by \citet{bai2022joint}. In this paper, the authors obtain a $1/(2+\varepsilon)$ approximation of the Offline surrogate by showing that $\OFFM(x)$ is monotone and submodular in $x$. Then, $\max_{x\in\X} \OFFM(x)$ corresponds to maximizing a monotone, submodular function subject to two partition constraints, for which \citet{lee2010maximizing} offers a $1/(2+\varepsilon)$-approximation algorithm for any $\varepsilon>0$.

% \bnote{Not deleting this in case we further develop this section.}

% To show this for a fixed realization of $D$ and SKU $\ell$, we randomly construct a feasible solution for $\OFF(R(x),D^\ell)$ that satisfies this last inequality in expectation. For the analysis, we will split the demand of each type $j$ into $D_{j\ell}$ identical types with one unit of demand each. After doing so, we end up with the same problem except that now $D_{j\ell}=1$ for all $j\in[T^\ell]$ under the new indexing, where $T^\ell$ is the total demand of SKU $\ell$ during the time horizon. (In particular, $\sum_{\ell \in [s]} T^\ell = T$.) Hereafter, we fix a demand realization $D$, an SKU $\ell$, split $D$ into unit demand nodes, and let $(y_{ij}^\ell)_{(i,j)\in [n]\times [T^\ell]}$ refer to an optimal solution to the LP defining $\OFF(x,D^\ell)$ {(after the splitting of types so that $y_{ij}^\ell\in[0,1]$ for all $i,j$)}.

\section{Experiments with Synthetic Data \label{sec:synthexp}}

{Generally speaking, an instance of the e-commerce placement and fulfillment problem consists of a quantity $Q$ to be placed across warehouses, a network defined by rewards $r_{ij}$ between each warehouse $i$ and demand location $j$, and a distribution over arrival sequences from different demand locations.
In \Cref{sec:techinical} we analyzed worst-case instances for network structure and the distribution over arrival sequences.
In this \namecref{sec:synthexp} we explore how our proposed methods perform when deployed on network structures and demand models that are common in the literature.
Computationally, we corroborate the intuition behind \Cref{thm:main}: choosing an inventory placement by optimizing the Offline surrogate leads to a good overall performance if a high-quality fulfillment policy is deployed downstream. Indeed, Offline Placement consistently performs among the highest rewards when paired with such a fulfillment policy. 

In order to answer this question, we evaluate the performance of four different inventory placement procedures together with five fulfillment policies on a range of 90 instances. Consistent with \citet{devalve2023understanding}, we find that a policy based on the shadow prices of the Offline surrogate for quantifying opportunity costs, and that periodically recomputes these shadow prices based on current inventories and observed demands, consistently outperforms all of the benchmarks. We compare the reward collected by this high-quality fulfillment policy when using the four different inventory placements and establish that the Offline Placement has the most robust performance.

The rest of this section is structured as follows. In \Cref{subsec:exp_procedures_policies}, we describe the different placement and fulfillment procedures we evaluate, and in \Cref{subsec:exp instances}, we describe all the instances we use to evaluate them. In \Cref{subsec:exp_procedure}, we outline the steps we follow to conduct our experiments. In \Cref{subsec:exp_results_policies} and \Cref{subsec:exp_results_placements}, we report the obtained results for comparing fulfillment policies and placement procedures, respectively.

\subsection{Placement Procedures and Fulfillment Policies \label{subsec:exp_procedures_policies}}

We evaluate four inventory placements by optimizing different surrogates, {and couple them with five different fulfillment procedures.} For a detailed description of these procedures, we refer the reader to \Cref{app:placement_desc} {and \Cref{app:fulfillment_desc}}.

\noindent\textbf{Offline Placement.} This placement procedure optimizes the Offline surrogate by solving its LP relaxation and rounding the solution. (We choose this way of solving the surrogate based on the findings in \Cref{app:lpvsgreedy}, where the LP rounding solution has the same performance as the greedy solution, but it is orders of magnitude faster to obtain.) Since the support of demand realizations is exponentially sized for all demand models, we solve the sample-average approximation (SAA) instead.  Remarkably, all the obtained solutions were integral, so no rounding was required.

\noindent\textbf{Myopic Placement.} This placement procedure optimizes the Myopic surrogate: the expected reward collected by a myopic fulfillment policy that chooses to deliver from the warehouse with remaining inventory that provides the highest immediate reward. This is optimized through simulation, using a greedy algorithm, and with SAA.

\noindent\textbf{Fluid Placement.} This placement procedure optimizes the Fluid surrogate, attempting to solve $\max_{x\in \X} \FLU(x)$. This is done by solving the linear relaxation and greedily rounding the obtained solution. Since we know the expected values for the aggregate demands, this procedure does not require sampling.

\noindent\textbf{Scaled Demand Fluid Placement.} This placement procedure optimizes a modified Fluid surrogate. In this variant, we scale the expected demand from each demand node so that the total expected demand is equal to the available inventory $Q$. This is intended as a fix to Fluid Placement when $Q$ exceeds the expected demand, leading to degenerate solutions.

\noindent\textbf{Fulfillment Policies.} We evaluate five different fulfillment policies. The first one is a simple \textit{myopic policy} that maximizes the immediate collected reward, without accounting for future demand. We also study two different fulfillment policies that account for the opportunity costs of depleting inventory units in certain warehouses by penalizing the collected rewards. We use the shadow prices of the inventory constraints of two different LPs as these penalties. Specifically, we use the Fluid LP and the Offline LP, leading to the \textit{fluid shadow price policy} (F-SP) and \textit{offline shadow price policy} (O-SP), respectively. The three policies described above define, for each demand node, a preference order of warehouses from which to deliver. This order is fixed across the time horizon, which could be suboptimal given particular realizations of demand samples. To improve this, we also evaluate the re-solving variants of these shadow price policies, where the shadow prices are recomputed taking into account the observed demand {(because we consider non-independent demand models, Bayesian updating is needed)} and current inventories. We call refer to these as F-SP-R and O-SP-R, for the Fluid and Offline LPs, respectively. To decide when to re-solve, we simulate arrival times associated to arrivals (explained in \Cref{subsubsec:demandmodelsexp}). In our experiments, we include two re-solving epochs: one after we observe the last arrival with an arrival time of $1/3$, and a second one after we observe the last arrival with an arrival time of $2/3$.

\subsection{Description of Instances \label{subsec:exp instances}}

Below is a description of the different instances we use in our experiments. A total of 90 instances arise from combinations of 3 network structures, 6 demand models, and 5 values for the total inventory $Q$.

\subsubsection{Description of Networks.}
Recall that a fulfillment network is described by a set of warehouses, a set of demand nodes, and the rewards between them.
% each pair of warehouses and demand nodes.

\noindent\textbf{Long-chain Network.} This follows the structure introduced by \citet{jordan1995principles}, with $n=m=5$ warehouses and demand nodes. Warehouse $i \in \{1,\dots,4\}$ connects to demand nodes $j=i$ and $j=i+1$, while warehouse $i=5$ connects to $j=5$ and $j=1$. Rewards for these edges are sampled from Uniform$(0,1)$, with all other pairs set to 0. This low-flexibility network allows each demand node to be served by at most two warehouses ($d=2$).

% \noindent\textbf{Long-chain Network.} This corresponds to the celebrated graph structure introduced \citet{jordan1995principles}. Here we have $n=m=5$ warehouses and demand nodes. Warehouse $i\in\{1,\dots,4\}$ can reach demand nodes $j=i$ and $j=i+1$. Warehouse $i=5$ that can reach $j=5$ and $j=1$. For all of these edges, the reward is fixed by sampling from a Uniform$(0,1)$ random variable, and for the rest of the warehouse and demand node pairs we set the rewards to 0. This is considered a low-flexibility network since every demand node can be served by at most two warehouses ($d=2$.)

\noindent\textbf{RDC-FDC Network.} Inspired by \citet{devalve2023understanding}, this network models JD.com’s operations with $n=m=5$ warehouses and demand nodes. Warehouse $i=1$ is the Regional Distribution Center (RDC), while $i=2,\dots,5$ are Front Distribution Centers (FDCs). Each FDC serves only its local demand node ($j=i$) with a reward of 1. The RDC can fulfill any demand: locally ($j=1$) with a reward of 1, and to $j=2,\dots,5$ with rewards sampled from Uniform$(0,1)$. This low-flexibility network allows each demand node at most two warehouses ($d=2$).

\noindent\textbf{Complete Network.} This graph consists of $n=5$ warehouses and $m=15$ demand nodes, randomly placed in $[0,1]^2$ and $[0.2,0.8]^2$, respectively. Every demand node is reachable from any warehouse, with rewards based on their Euclidean distance: $r_{ij} = 1 - d_{ij}/\sqrt{2}$ for all $(i,j) \in [n] \times [m]$.

% \noindent\textbf{Complete Network.} This graph has $n=5$ warehouses and $m=15$ demand nodes. Demand nodes are randomly placed within the $[0,1]^2$ square. Warehouses are randomly placed within the $[0.2,0.8]^2$ square. As the name suggests, all demand nodes are reachable from all warehouses, and the reward is determined by how far a warehouse is from the demand node. Specifically, for all $(i,j)\in[n]\times[m]$ we set $r_{ij}= 1-d_{ij}/\sqrt{2}$, where $d_{ij}$ is the Euclidean distance between warehouse $i$ and demand node $j$.

\subsubsection{Description of Demand Models. \label{subsubsec:demandmodelsexp}} A demand model specifies a distribution over demand arrival sequences. We consider three models: the classic temporal independence model, its extension with a random time horizon, and a spatial independence variant with a random arrival order. Each demand node $j$ has a weight $p_j \in [0,1]$ representing its average arrival proportion, with $\sum_{j\in[m]}p_j=1$. We define two weight-setting methods, yielding six stochastic arrival models.

\noindent\textbf{Deterministic-Horizon Temporal Independence ($\DHTI$).} Here, demand always consists of $T=60$ arrivals, each assigned to demand node $j$ with probability $p_j$. This IID model is well-studied in revenue management and online matching \citep{gallego1994optimal, feldman2009online}. 

\noindent\textbf{Random-Horizon Temporal Independence ($\RHTI$).} This model extends $\DHTI$ by making $T$ random \citep{ aouad2022nonparametric}. Here, we sample $T$ from a Geometric$(q)$ distribution with $q=1/61$, ensuring $\E[T]=60$. Each arrival is then assigned to demand node $j$ with probability $p_j$. This high-variance distribution has a standard deviation roughly equal to its mean.  

\noindent\textbf{Random-Order Spatial Independence ($\ROSI$).} Here, total demand per node $D_j$ follows a Geometric$(q_j)$ distribution, ensuring $\E[D_j] = 60p_j$ and $\E[T] = 60$. Once sampled, arrivals occur in random order. This modifies the Indep model from \citet{aouad2022nonparametric} by introducing randomness instead of an adversarial order.

We define two weight-setting methods: \textit{uniform weights}, where $p_j = 1/m$, and \textit{reward-based weights}, where $p_j$ is proportional to the total reward from warehouse-to-node edges. The latter increases demand near warehouses, aligning with real-world facility location strategies.

For $\RHTI$ and $\ROSI$, we also simulate \textit{arrival times} for fulfillment policies. Given $T$ arrivals, we sample and sort $T$ Uniform$(0,1)$ variables as arrival times. These inform better decisions; e.g., if early arrivals exceed expectations, later demand is likely higher as well.

\subsubsection{Initial Inventory.} We choose the total inventory $Q$ to take the values in $\{30, 45, 60, 75, 90\}$. Since the expected number of demand arrivals is 60 in all demand models, these initial inventories are such that the ratio between the available inventory and the expected total demand lies within $\{0.50, 0.75, 1.00, 1.25, 1.50\}$.

\subsection{Experimental Setup \label{subsec:exp_procedure}} To answer our questions, we execute the following experiment. For each of the 90 instances, we:
\begin{enumerate}
    \item \textbf{Inventory placements:} Compute four placements. Fluid placements use exact demand distributions, while Offline and Myopic placements rely on 1000 sampled demand sequences to solve their SAA versions.
    
  \item \textbf{Fulfillment policies:} Construct priority lists for each demand node, fulfillment algorithm (myopic, fluid, offline), and starting inventory. Non-myopic policies depend on initial inventory since different levels lead to varying shadow prices. The Offline LP uses the same 1000 demand samples as in step 1.

    \item \textbf{Performance simulation:} Generate 1000 fresh demand sequences and compute the average reward for each policy across all 20 combinations of 4 inventory placements and 5 fulfillment policies.
\end{enumerate}
We normalize collected rewards by the \textit{prophet's reward}, defining the \textit{competitive ratio} as this ratio. The prophet's reward is the linear relaxation $\max_{x\in\CH(\X)} \hOFF(x)$, computed using the 1000 demand samples from step 3. It represents the reward of an offline-fulfilling prophet who knows the empirical demand distribution. This value, independent of inventory placement, serves as an upper bound on expected rewards, ensuring the competitive ratio lies between 0 and 1.
% In each instance, we normalize the reward collected by inventory placement and fulfillment policy pairs by what we call the \textit{prophet's reward}, and we refer to this ratio as the \textit{competitive ratio} of a solution pair. The prophet's reward corresponds to the value of the linear relaxation $\max_{x\in\CH(\X)} \hOFF(x)$ when we feed the 1000 demand samples used in step 3. It can be interpreted as the reward collected by a prophet who can perform offline fulfillment and also knows the empirical distribution induced by the demand samples used to test the performance of solution pairs. (The prophet is also allowed to make fractional inventory placement decisions, although this did not happen in our experiments.) This value only depends on the instance, not the inventory placement.
% It is also an upper bound to whatever expected reward we can hope to obtain with any solution pair, and thus the competitive ratio will be a value between 0 and 1.

\subsection{Results: Comparison Between Fulfillment Policies \label{subsec:exp_results_policies}} We start by comparing the performance obtained by deploying different fulfillment policies combined with different inventory placements.

As an aggregate measurement of performance, \Cref{tab:fulfillment} shows the competitive ratio obtained by each pair of solutions, averaged over the 90 instances. The main observation from this table is that the offline shadow prices policy with re-solving achieves the best average competitive ratio for all inventory placements. This dominance goes beyond the average: this fulfillment policy achieves the best performance in 310 out of the 360 pairs of instances and inventory placement procedures. Moreover, even when it is not the best policy, its collected reward is always within 0.5\% of the best policy for any instance and inventory placement pair. This is consistent with the findings in \citet{devalve2023understanding}, where they also find that this particular policy performs the best among other practically motivated policies.

\begin{table}[h]
    \centering
    \begin{tabular}{l S S S S}
    \toprule
     & {\shortstack{Offline\\Placement}} & {\shortstack{Fluid\\Placement}} & {\shortstack{Scaled Demand\\Fluid Placement}} & {\shortstack{Myopic\\Placement}} \\
    \midrule
    Myopic & 0.956 & 0.933 & 0.960 & 0.966 \\
    F-SP   & 0.936 & 0.870 & 0.939 & 0.943 \\
    O-SP   & 0.974 & 0.944 & 0.965 & 0.969 \\
    F-SP-R & 0.957 & 0.912 & 0.955 & 0.960 \\
    O-SP-R & 0.986 & 0.954 & 0.974 & 0.978 \\
    \bottomrule
    \end{tabular}
    \caption{Average competitive ratio obtained by each pair of placement procedure and fulfillment policy.}
    \label{tab:fulfillment}
\end{table}

We can also observe that for each fulfillment policy, the best performance is obtained with either the Offline Placement (O-SP, O-SP-R) or the Myopic Placement (Myopic, F-SP, F-SP-R). This suggests that inventory placements that account for randomness in the demand tend to outperform placement procedures with a certainty-equivalence approach.

% For a more fine-grained analysis of fulfillment policy performance, we refer the reader to \bnote{Appendinx that might or might not exist.}

\subsection{Results: Comparison Between Placement Procedures \label{subsec:exp_results_placements}}

We now turn our focus to a more fine-grained comparison between inventory placements. Since we established that the offline shadow prices policy with re-solving is best-performing, we hereafter focus on results under it, to reflect a high-quality fulfillment policy.

\Cref{fig:competitiveratiouni} shows the competitive ratio for the 45 instances with uniform weights. (The plots for reward-based weights are presented in \Cref{app:rw_weights}. The shapes of the graphs and the insights remain unchanged.) The main conclusion drawn from these plots is the superior robustness of Offline Placement. It can be seen that this placement procedure obtains the best or nearly the best performance across all of the displayed instances. Moreover, there is no other procedure that is always close to Offline Placement. For instance, in the long-chain network, the only procedure that is close to Offline Placement for $Q\leq 60$ is the Fluid Placement. On the other hand, if $Q\geq 60$, Fluid Placement has a far worse performance, but Myopic Placement and Scaled Demand Fluid Placement become competitive. 

\begin{figure}[h]
    \centering
    \hspace*{-0.5cm}
    \begin{tikzpicture}
    % Begin a groupplot with shared legend options.
    \begin{groupplot}[
        group style={
            group size=3 by 3,
            vertical sep=1.2cm,
            horizontal sep=0.8cm,
            ylabels at=edge left,   % Place y-axis labels at the left edge
            xlabels at=edge bottom  % Place x-axis labels at the bottom edge
        },
        height=5cm, width=6cm,
        title style={yshift=1.2cm, font=\large},  % Shift column titles upward
        every axis y label/.style={
            at={(axis description cs:-0.3,0.5)},
            anchor=south,
            rotate=90
        },
        every axis title/.style={
            at={(axis description cs:0.5,1.2)},
            anchor=south
        },
        legend to name=globalLegend,   % This collects legend entries under the name "globalLegend"
        legend columns=4               % Arrange the legend entries in one row
    ]
    
    % --- First Row of Plots ---
    % First subplot (the one where we add legend entries)
    \nextgroupplot[title={\shortstack{Deterministic-Horizon\\Temporal Independence}}, ylabel={Long-chain}, ymin=0.89, ymax=1.005]
        \addplot table [x=total_inventory, y=offline_lp_rounding, col sep=comma] {uniform_longchain_deter.csv};
        \addlegendentry{Offline LP rounding}
        \addplot table [x=total_inventory, y=fluid_lp_rounding, col sep=comma] {uniform_longchain_deter.csv};
        \addlegendentry{Fluid LP rounding}
        \addplot table [x=total_inventory, y=fluid_lp_rounding_withscaling, col sep=comma] {uniform_longchain_deter.csv};
        \addlegendentry{Fluid LP rounding with scaling}
        \addplot table [x=total_inventory, y=myopic_greedy, col sep=comma] {uniform_longchain_deter.csv};
        \addlegendentry{Myopic greedy}
    
    % Second subplot of the first row (no legend entries here)
    \nextgroupplot[title={\shortstack{Random-Horizon\\Temporal Independence}}, ymin=0.89, ymax=1.005, yticklabels={}]
        \addplot table [x=total_inventory, y=offline_lp_rounding, col sep=comma] {uniform_longchain_correl.csv};
        \addplot table [x=total_inventory, y=fluid_lp_rounding, col sep=comma] {uniform_longchain_correl.csv};
        \addplot table [x=total_inventory, y=fluid_lp_rounding_withscaling, col sep=comma] {uniform_longchain_correl.csv};
        \addplot table [x=total_inventory, y=myopic_greedy, col sep=comma] {uniform_longchain_correl.csv};
    
    % Third subplot of the first row
    \nextgroupplot[title={\shortstack{Spatial Independence}}, ymin=0.89, ymax=1.005, yticklabels={}]
        \addplot table [x=total_inventory, y=offline_lp_rounding, col sep=comma] {uniform_longchain_indep.csv};
        \addplot table [x=total_inventory, y=fluid_lp_rounding, col sep=comma] {uniform_longchain_indep.csv};
        \addplot table [x=total_inventory, y=fluid_lp_rounding_withscaling, col sep=comma] {uniform_longchain_indep.csv};
        \addplot table [x=total_inventory, y=myopic_greedy, col sep=comma] {uniform_longchain_indep.csv};

    % --- Second Row of Plots ---
ch    \nextgroupplot[ylabel={\shortstack{RDC-FDC}}, ymin=0.755, ymax=1.005]
        \addplot table [x=total_inventory, y=offline_lp_rounding, col sep=comma] {uniform_starlike_deter.csv};
        \addplot table [x=total_inventory, y=fluid_lp_rounding, col sep=comma] {uniform_starlike_deter.csv};
        \addplot table [x=total_inventory, y=fluid_lp_rounding_withscaling, col sep=comma] {uniform_starlike_deter.csv};
        \addplot table [x=total_inventory, y=myopic_greedy, col sep=comma] {uniform_starlike_deter.csv};

    \nextgroupplot[ymin=0.755, ymax=1.005, yticklabels={}]
        \addplot table [x=total_inventory, y=offline_lp_rounding, col sep=comma] {uniform_starlike_correl.csv};
        \addplot table [x=total_inventory, y=fluid_lp_rounding, col sep=comma] {uniform_starlike_correl.csv};
        \addplot table [x=total_inventory, y=fluid_lp_rounding_withscaling, col sep=comma] {uniform_starlike_correl.csv};
        \addplot table [x=total_inventory, y=myopic_greedy, col sep=comma] {uniform_starlike_correl.csv};

    \nextgroupplot[ymin=0.755, ymax=1.005, yticklabels={}]
        \addplot table [x=total_inventory, y=offline_lp_rounding, col sep=comma] {uniform_starlike_indep.csv};
        \addplot table [x=total_inventory, y=fluid_lp_rounding, col sep=comma] {uniform_starlike_indep.csv};
        \addplot table [x=total_inventory, y=fluid_lp_rounding_withscaling, col sep=comma] {uniform_starlike_indep.csv};
        \addplot table [x=total_inventory, y=myopic_greedy, col sep=comma] {uniform_starlike_indep.csv};
    
    % --- Third Row of Plots ---
    \nextgroupplot[ylabel={Complete}, xlabel={Total inventory}, ymin=0.935, ymax=1.005]
        \addplot table [x=total_inventory, y=offline_lp_rounding, col sep=comma] {uniform_complete_deter.csv};
        \addplot table [x=total_inventory, y=fluid_lp_rounding, col sep=comma] {uniform_complete_deter.csv};
        \addplot table [x=total_inventory, y=fluid_lp_rounding_withscaling, col sep=comma] {uniform_complete_deter.csv};
        \addplot table [x=total_inventory, y=myopic_greedy, col sep=comma] {uniform_complete_deter.csv};

    \nextgroupplot[xlabel={Total inventory}, ymin=0.935, ymax=1.005, yticklabels={}]
        \addplot table [x=total_inventory, y=offline_lp_rounding, col sep=comma] {uniform_complete_correl.csv};
        \addplot table [x=total_inventory, y=fluid_lp_rounding, col sep=comma] {uniform_complete_correl.csv};
        \addplot table [x=total_inventory, y=fluid_lp_rounding_withscaling, col sep=comma] {uniform_complete_correl.csv};
        \addplot table [x=total_inventory, y=myopic_greedy, col sep=comma] {uniform_complete_correl.csv};

    \nextgroupplot[xlabel={Total inventory}, ymin=0.935, ymax=1.005, yticklabels={}]
        \addplot table [x=total_inventory, y=offline_lp_rounding, col sep=comma] {uniform_complete_indep.csv};
        \addplot table [x=total_inventory, y=fluid_lp_rounding, col sep=comma] {uniform_complete_indep.csv};
        \addplot table [x=total_inventory, y=fluid_lp_rounding_withscaling, col sep=comma] {uniform_complete_indep.csv};
        \addplot table [x=total_inventory, y=myopic_greedy, col sep=comma] {uniform_complete_indep.csv};
    
    \end{groupplot}
    \end{tikzpicture}
    
    % Now place the shared legend (adjust its placement as desired)

    % Now place the shared legend by referencing its name
    \vspace{0.5cm} % Adjust spacing as needed
  \begin{tikzpicture}[scale=1]
  % Offline LP rounding (blue, with a circle marker)
  \draw[myBlue, thick, dashed,
    decoration={
      markings,
      mark=at position 0.5 with {
        \node[fill=myBlue, draw=myBlue, solid, shape=circle, inner sep=2pt] {};
      }
    },
    postaction={decorate}
  ] (0,0) -- (1,0);
  \node[right] at (1.2,0) {Offline Placement};

  % Fluid LP rounding (orange, with a solid square marker)
  \draw[myOrange, thick, dashed,
    decoration={
      markings,
      mark=at position 0.5 with {
        \node[fill=myOrange, draw=myOrange, solid, shape=rectangle, inner sep=2.5pt] {};
      }
    },
    postaction={decorate}
  ] (0,-0.5) -- (1,-0.5);
  \node[right] at (1.2,-0.5) {Fluid Placement};

  % Fluid LP rounding with scaling (green, with a triangle marker)
  \draw[myGreen, thick, dashed,
    decoration={
      markings,
      mark=at position 0.5 with {
        \node[fill=myGreen, draw=myGreen, solid, 
          shape=regular polygon, regular polygon sides=3, inner sep=1.3pt, rotate=120] {};
      }
    },
    postaction={decorate}
  ] (0,-1) -- (1,-1);
  \node[right] at (1.2,-1) {Scaled Demand Fluid Placement};

  % Myopic greedy (magenta, with a diamond marker)
  \draw[myMagenta, thick, dashed,
    decoration={
      markings,
      mark=at position 0.5 with {
        \node[fill=myMagenta, draw=myMagenta, solid, shape=diamond, inner sep=1.8pt] {};
      }
    },
    postaction={decorate}
  ] (0,-1.5) -- (1,-1.5);
  \node[right] at (1.2,-1.5) {Myopic Placement};
\end{tikzpicture}

\caption{Competitive ratio as a function of total inventory, for all combinations of graphs and demand models with uniform weights. Each row corresponds to a different network structure and each column corresponds to a different demand model.}
\label{fig:competitiveratiouni}
    
\end{figure}

Another important observation is that placement procedures that account for randomness in demand (Offline and Myopic) generally outperform the certainty equivalence counterparts (Fluid and Scaled Demand Fluid). The only exception occurs in the long-chain network with $Q< 60$. 
% This is in line with the findings in \citet{devalve2023understanding}.

One last observation is that Fluid Placement seems to be especially brittle. This is expected for instances where $Q>60$ as the total inventory is greater than the total expected demand, which leads to ``leftover'' inventory. This leftover inventory can be placed arbitrarily and still obtain an optimal solution, so in these cases the solution to the Fluid LP becomes degenerate. This can be fixed for $Q>60$ by scaling up the demand so it is equal to the total inventory, but this comes at a cost of potential low performance when $Q<60$, as observed in the long-chain network. A mixed procedure that scales the demand if $Q>60$ and does not if $Q\leq 60$ also fails to obtain the robustness seen for Offline Placement. This can be observed in the RDC-FDC network, where Fluid Placement performs worse than its Scaled Demand counterpart even if $Q< 60$.

In broader terms, the complete graph seems to be the hardest instance in terms of maximizing the competitive ratio. In the long-chain and RDC-FDC networks, Offline Placement achieves a competitive ratio of at least 0.98, regardless of the starting inventory. For the complete network, this ratio can be as low as 0.94 for low inventory regimes. This could be due to the high flexibility of the complete network, where more subtle hindsight optimal assignments can be constructed. In terms of demand models, $\RHTI$ and $\ROSI$ appear to be harder than $\DHTI$. Which demand model is the hardest to approximate depends on the graph structure.

\

\textbf{Acknowledgments.} The authors sincerely thank anonymous reviewers for identifying an error in the original proof of \Cref{lem:weightsExist}.  The statement is correct and the proof has since been fixed.

\bibliographystyle{informs2014} % outcomment this and next line in Case 1
\bibliography{main} % if more than one, comma separated

\clearpage

% Appendix here
% Options are (1) APPENDIX (with or without general title) or
%             (2) APPENDICES (if it has more than one unrelated sections)
% Outcomment the appropriate case if necessary
%
% \begin{APPENDIX}{<Title of the Appendix>}
% \end{APPENDIX}
%
%   or
%

\begin{APPENDICES}
\crefalias{section}{appendix}
\crefalias{subsection}{appendix}

\linespread{\linespace}\selectfont{}
\section{Proofs of \Cref{sec:techinical} and \Cref{sec:joinproblem}}

\subsection{Proof of \Cref{lem:weightsExist} \label{app:proof_weights_lemma}}
For brevity, we remove index $i$ from the proof. In particular, we will use $x$ to denote the LP variable for inventory in the $i$-th warehouse, rather than the complete vector.
 Before proving \Cref{lem:weightsExist}, we prove the following auxiliary claim.

\begin{claim}\label{claim:1}
\[ \sum_{j\in[T]} \max\left\{0,\frac{y_{j}-x^f}{1-x^f}\right\} \leq \lfloor x \rfloor. \]
\end{claim}
\proof{Proof.} Let $S:=\{ j : y_{j} \geq x^f\}$, $V := \sum_{j\in S} y_{j}$, and $V^f := V-\lfloor V \rfloor $. We have
\begin{equation*}
    \ \sum_{j\in[T]} \max\left\{0,\frac{y_{j}-x^f}{1-x^f}\right\}   = \sum_{j\in S} \frac{y_{j}-x^f}{1-x^f} = \frac{V^f + \lfloor V \rfloor - |S|x^f}{1-x^f} \leq \frac{V^f + \lfloor V \rfloor - \lceil V \rceil x^f}{1-x^f}, \quad (\ddagger)
\end{equation*}
since $|S| \geq \lceil V \rceil$. We distinguish 2 cases:

\underline{Case 1: $V^f=0$}. In this case, $\lceil V \rceil = \lfloor V \rfloor$, so the right-hand side of $(\ddagger)$ is equal to $\lfloor V \rfloor \leq \lfloor x \rfloor$. 

\underline{Case 2: $V^f>0$}. In this case, $\lceil V \rceil = \lfloor V \rfloor + 1$ and the right-hand side of $(\ddagger)$ is equal to
\[ \lfloor V \rfloor + \frac{V^f - x^f}{1-x^f}. \]
We further distinguish two subcases.

\underline{Subcase 2.1: $V\geq \lfloor x \rfloor$}. Here, $V^f \leq x^f$ because $V\leq x$. Therefore, the right-hand side of $(\ddagger)$ is at most $\lfloor V \rfloor \leq \lfloor x \rfloor$.

\underline{Subcase 2.2: $ V< \lfloor x \rfloor$}. In this case, $\lfloor V \rfloor \leq \lfloor x \rfloor -1$, so the right-hand side of $(\ddagger)$ is at most
\[ \lfloor x \rfloor -1 + \frac{1-x^f}{1-x^f} = \lfloor x \rfloor. \qed \]

\proof{Proof of \Cref{lem:weightsExist}.} The result is trivial if $\sum_{j\in[T]} y_{j}^\samplePathOld \leq \lfloor x \rfloor$. Indeed, we can simply set $y_{j}^{\samplePathOld\LL} = y_{j}^{\samplePathOld\HH} = y_{j}^{\samplePathOld}$ for all $j\in [T]$. Therefore, assume $\sum_{j\in[T]} y_{j}^\samplePathOld > \lfloor x \rfloor$. We propose the following solution: for all $j\in[T]$ set
\[y_{j}^{\mathrm{L}} = \max\left\{ \frac{y_{j} -x^f}{1-x^f} ,0, y_{j} - \lambda^*\right\} \text{ and } y_{j}^{\mathrm{H}} = \frac{y_{j} - (1-x^{f})y_{j}^{\mathrm{L}}}{x^f},\]
where $\lambda^*\in[0,1]$ is such that $\sum_{j\in[T]}y_{j}^\mathrm{L} = \lfloor x \rfloor$.
We first show the existence of such $\lambda^*$. Define  
\[\Delta(\lambda):=\sum_{j\in[T]}y_{j}^\mathrm{L} = \sum_{j\in[T]} \max\left\{ \frac{y_{j} -x^f}{1-x^f} ,0, y_{j} - \lambda\right\}.\]
Clearly $\Delta(\lambda)$ is a continuous {decreasing} function. {We have by assumption that}
\[ \Delta(0) \geq \sum_{j\in[T]} y_{j} > \lfloor x \rfloor. \]
On the other hand, by \Cref{claim:1} it holds that
\[ \Delta(1) = \sum_{j\in[T]} \max\left\{0,\frac{y_{j}-x^f}{1-x^f}\right\} \leq \lfloor x \rfloor. \]
Therefore, the Intermediate Value Theorem says that there exists $\lambda^*\in[0,1]$ such that $\Delta(\lambda^*) = \lfloor x \rfloor$.

We now show that our proposed solution satisfies \Cref{eq:marginal_dist_y,eq:neg_corr_y,eq:inv_constraint1,eq:inv_constraint2}. Equations \Cref{eq:marginal_dist_y,eq:inv_constraint1} are satisfied by definition. To see that \Cref{eq:inv_constraint2} is verified, see that
\begin{align*}
    \sum_{j\in[T]} y_{j}^\mathrm{H} &= \frac{\sum_{j\in[T]} y_{j} -(1-x^f)\sum_{j\in[T]}y_{j}^\mathrm{L} }{x^f}
= \frac{\sum_{j\in[T]} y_{j} -(1-x^f)\lfloor x \rfloor }{x^f}\\
& \leq \frac{x-(1-x^f)\lfloor x \rfloor }{x^f}
 =  \frac{x^f +\lfloor x \rfloor -(1-x^f)\lfloor x \rfloor }{x^f} = 1 + \lfloor x \rfloor.
\end{align*}
To see that \Cref{eq:neg_corr_y} is satisfied, first notice that $y_{j}^\mathrm{L}\geq 0 $ for all $j\in[T]$. We also have that $y_{j}^\mathrm{L} \leq y_{j}$, which combined with \Cref{eq:marginal_dist_y} implies that $y_{j}^\mathrm{L} \leq y_{j}^\mathrm{H}$. Finally, for all $j\in[T]$ we have
\[y_{j}^\mathrm{H} = \frac{y_{j} - (1-x^{f})y_{j}^{\mathrm{L}}}{x^f} \leq \frac{y_{j} - (y_{j}-x^f)}{x^f} =1. \qed\]

\subsection{Proof of \Cref{lem:ZBound} \label{app:proof_Zbound}}

Before proving \Cref{lem:ZBound}, we prove the following claim. This allows to transfer the negative correlation property of $(W_i)_{i\in[n]}$ across $i$ to the preliminary assignments $(Y_{ij})_{i\in[n]}$, for each demand node $j\in[T]$.
\begin{claim}\label{lem:neg_corr}
    For any $j\in[T]$ and for any subset $S\subseteq [n]$, it holds that
    \[ \PP\left( \bigcap_{i\in S} Y_{ij}^\samplePathOld = 0 \right) \leq   \prod_{i\in S} \PP(Y_{ij}^\samplePathOld = 0). \]
\end{claim}
\proof{Proof.} Fix $j\in [T]$. We can write
\[Y_{ij}^{\samplePathOld} = W_i Y_{ij}^\HH +  (1-W_i)Y_{ij}^\LL = Y_{ij}^\HH - (1-W_i)(Y_{ij}^\HH - Y_{ij}^\LL),\]
where $Y_{ij}^{\samplePathOld \HH}$ and $Y_{ij}^{\samplePathOld \LL}$ are Bernoulli random variables with means $y_{ij}^{\samplePathOld \HH}$ and $y_{ij}^{\samplePathOld \LL}$, respectively, independent across $i$, and coupled such that $Y_{ij}^{\samplePathOld \LL}\leq Y_{ij}^{\samplePathOld \HH}$ with probability 1 (we can do this because $y_{ij}^{\samplePathOld \LL} \leq y_{ij}^{\samplePathOld \HH}$). Hence,
\begin{align}
\PP\left( \bigcap_{i\in S} Y_{ij}^\samplePathOld = 0 \right) &=\bE\left[\prod_{i\in S} (1-Y_{ij}^{\samplePathOld})\right]\nonumber\\
&= \bE\left[\prod_{i\in S} (1-Y^{{\samplePathOld}\HH}_{ij}+(1-W_i)(Y^{\samplePathOld\HH}_{ij} - Y^{\samplePathOld\LL}_{ij}))\right]\nonumber\\ 
&= \sum_{T\subseteq S}\bE\left[\prod_{i\in T}(1-Y^{\samplePathOld\HH}_{ij})\prod_{i\in S\setminus T}(1-W_i)(Y^{\samplePathOld\HH}_{ij} - Y^{\samplePathOld\LL}_{ij})\right]\nonumber\\ 
&= \sum_{T\subseteq S}\prod_{i\in T}(1-y^{\samplePathOld\HH}_{ij})\prod_{i\in S\setminus T}(y^{\samplePathOld\HH}_{ij} - y^{\samplePathOld\LL}_{ij})\bE\left[\prod_{i\in S\setminus T}(1-W_i)\right]\label{eq:independence}\\ 
&\le\sum_{T\subseteq S}\prod_{i\in T}(1-y^{\samplePathOld\HH}_{ij})\prod_{i\in S\setminus T}(y^{\samplePathOld\HH}_{ij} - y^{\samplePathOld\LL}_{ij})(1-x_i^f)\label{eq:neg_correlation_x}\\ 
&=\prod_{i\in S} (1-y^{\samplePathOld\HH}_{ij}+(1-x_i^f)(y^{\samplePathOld\HH}_{ij} - y^{\samplePathOld\LL}_{ij}))\nonumber\\
&= \prod_{i\in S} (1- y_{ij}^\samplePathOld)\nonumber\\
& =  \prod_{i\in S} \PP( Y_{ij}^\samplePathOld = 0)\nonumber.
\end{align}
In \Cref{eq:independence} we use that $Y_{ij}^{\samplePathOld \HH},Y_{ij}^{\samplePathOld \LL}$ are independent across $i$, and that for each term in the sum we are multiplying $(1-Y_{ij}^{\samplePathOld \HH})$ and $Y_{ij}^{\samplePathOld \HH} - Y_{ij}^{\samplePathOld \LL}$ on disjoint sets. In Inequality (\ref{eq:neg_correlation_x}) we use the negative correlation property (P3) of the dependent rounding scheme by \citet{gandhi2006dependent}.\qed

\proof{Proof of \Cref{lem:ZBound}.}
Fix $j$ and re-label the warehouses so that $r_{1j} \geq r_{2j} \geq \cdots \geq r_{d_j j}{\ge r_{d_j+1,j}=\cdots=r_n=0}$. Since the unit of demand of node $j$ will be assigned to the warehouse that offers the highest reward, we can write $Z_{ij}^\samplePathOld = Y_{ij}^\samplePathOld \prod_{i'< i} (1-Y_{i'j}^\samplePathOld)$ for all $i=1,\ldots,d_j$. In words, to assign $j$ to $i$ we need $Y_{ij}^\samplePathOld=1$ and $Y_{i'j}^\samplePathOld=0$ for all $i'$ with higher reward than $i$. Thus,
\begin{align}
     \E\left[\sum_{i\in[n]} r_{ij}Z_{ij}^\samplePathOld \right] & =  \E\left[\sum_{i=1}^{d_j} r_{ij}  Y_{ij}^\samplePathOld \prod_{i'< i} (1-Y_{i'j}^\samplePathOld)\right]\nonumber \\
    & = \sum_{i=1}^{d_j}(r_{i,j}-{r_{i+1,j}})\left(1 - \PP\left(\bigcap_{i'\leq i} Y_{i'j}^{\samplePathOld}=0\right)\right)\label{eq:incremental}\\
    &\geq \sum_{i=1}^{d_j}(r_{i,j}-{r_{i+1,j}})\left(1 - \prod_{i'\leq i}(1- y_{i'j}^{\samplePathOld})\right)\label{eq:negative_correlation_application}\\
    & \geq \sum_{i=1}^{d_j}(r_{i,j}-{r_{i+1,j}})\left(1 - \left(1-\frac{1}{i}\right)^i\right) \label{eq:moving_mass_inequality}\\
    & \geq r_{1j}\left(1 - \left(1-\frac{1}{d_j}\right)^{d_j}\right)\label{eq:telescope}\\
    & \geq \left(1 - \left(1-\frac{1}{d_j}\right)^{d_j}\right) \sum_{i\in[n]}r_{ij}y_{ij}^\samplePathOld.\label{eq:y_sum_1}
\end{align}
In \Cref{eq:incremental} we expressed the expected reward using increments, where we gain $r_{i,j}-r_{i+1,j}$ if and only if at least one of $Y_{i'j}^\samplePathOld$ for $i'\leq i$ turns out to be 1. In Inequality (\ref{eq:negative_correlation_application}) we use the negative correlation property from \Cref{lem:neg_corr}. In Inequality (\ref{eq:moving_mass_inequality}) we use $\sum_{i\in[n]} y_{ij}^\samplePathOld \leq 1$ (since $D_j=1$ for all $j$ with the new indexing), {from which it follows elementarily that} the product $\prod_{i'\leq i}(1- y_{i'j}^{\samplePathOld})$ is maximized by setting $y_{i'j}^\samplePathOld=1/i$ for $i'\leq i$ and 0 otherwise. {Inequality (\ref{eq:telescope}) follows by a telescoping sum after using the fact that expression $1-(1-1/i)^i$ is minimized when $i=d_j$.}
Finally, in Inequality (\ref{eq:y_sum_1}) we again use that $\sum_{i\in[n]} y_{ij}^\samplePathOld \leq 1$  and that $r_{1j}\geq r_{i'j}$ for all $i'\neq i$. \qed

\subsection{Proof of \Cref{thm:rand_rounding} \label{app:proof_rounding}}

% We have
% \begin{align}
%     \E_R[\OFF(R(x))] = \E\left[\sum_{\samplePathOld\in \Omega} q_\samplePathOld \OFF(R(x),D^\samplePathOld)\right] =\sum_{\samplePathOld\in \Omega} q_\samplePathOld \E_R\left[ \OFF(R(x),D^\samplePathOld)\right]. \label{eq:exp_offline_round}
% \end{align}
Recall we just need to prove on a fixed realization of $D$ that $\E_R[ \OFF(R(x),D^\samplePathOld) ] \geq (1 - \left(1-\frac{1}{d})^d\right)\OFF(x, D^\samplePathOld)$. We proceed by showing that, for any demand realization $D$, $(Z_{ij}^\samplePathOld)_{(i,j)\in [n]\times[T]}$ is a feasible solution for the LP defining $\OFF(R(x),D^\samplePathOld)$. Indeed, the degree preservation property (P2) gives us
\[ \sum_{j\in [T]} Z_{ij}^\samplePathOld \leq \sum_{j\in [T]} Y_{ij}^\samplePathOld \leq R_i(x) \]
for all $i\in[n]$, since the randomized rounding weights satisfy \Cref{eq:inv_constraint1,eq:inv_constraint2} so we will not exceed the inventory constraints either if inventory gets rounded up or down. We also have  $\sum_{i\in [n]} Z_{ij}^\samplePathOld \leq  D_{j}^\samplePathOld =1$ for all $j\in[T]$. Indeed, we have $Z_{ij}^\samplePathOld = Y_{ij}^\samplePathOld \prod_{i'\leq i} (1-Y_{i'j}^\samplePathOld)$ which will only be 1 for at most one $i\in[n]$ because we assign the demand node to the single warehouse with the highest reward. It follows that, with probability 1,
\[ \OFF(R(x),D^\samplePathOld) \geq \sum_{j\in[T]}\sum_{i\in[n]} r_{ij}Z_{ij}^\samplePathOld.\]
By taking expectations we get
\begin{align}
    \E_{R}[\OFF(R(x),D^\samplePathOld)] &= \E_{R}[\E_Z[\OFF(R(x),D^\samplePathOld)|R(x)]]  \nonumber \\
    &\geq \E_R \left[\E_Z\left[\sum_{j\in[m]}\sum_{i\in[n]} r_{ij}Z_{ij}^\samplePathOld \Bigg| R(x)\right]\right]\nonumber\\
    & = \sum_{j\in[T]}\sum_{i\in[n]} r_{ij} \E_Z[Z_{ij}^\samplePathOld]\label{eq:iter}\\
    & \geq  \sum_{j\in[T]}\left(1-\left(1-\frac{1}{d_j}\right)^{d_j}\right)\sum_{i\in[n]} r_{ij} y_{ij}^\samplePathOld\label{eq:degree_lemma}\\
     & \geq \left(1-\left(1-\frac{1}{d}\right)^{d}\right) \sum_{j\in[T]}\sum_{i\in[n]} r_{ij}  y_{ij}^\samplePathOld\label{eq:decreasing_d}\\
     & =  \left(1-\left(1-\frac{1}{d}\right)^{d}\right) \OFF(x,D^\samplePathOld),\qed\nonumber
\end{align}
where in \Cref{eq:iter} we use linearity of expectation and law of iterated expectations, in Inequality (\ref{eq:degree_lemma}) we use \Cref{lem:ZBound} and in Inequality (\ref{eq:decreasing_d}) we use that $ \left(1-\left(1-\frac{1}{d_j}\right)^{d_j}\right) $ is decreasing in $d_j$. \qed

\subsection{Proof of \Cref{lem:tight_example} \label{app:tight_example}}

Let $\{n_k\}_{k\geq 1}$ be an increasing sequence of natural numbers such that $n_k/d$ is an integer. Consider the following family of instances indexed by $k$, with $n_k$ warehouses and $\binom{n_k}{d}$ demand nodes, each one of them served by a different subset of $d$ warehouses. The rewards are all equal to 1, and the demand distribution is such that only one of the demand nodes will have demand equal to 1, and the remaining will have demand equal to 0. The total inventory to be distributed is $Q= n_k/d$. The optimal solution for $\max_{x\in \CH(\X)} \OFF(x)$ is to set $x_i = 1/d$ for all $i\in [n]$, yielding an optimal value of 1. On the other hand, an optimal integer placement places 1 unit of inventory in $n_k/d$ different warehouses. The reward collected by the offline algorithm is equal to 1 minus the probability of the realized demand node being one of the nodes that can be served given the inventory placement. That is,
\begin{align*}
    1 - \prod_{\ell=0}^{d-1} \frac{n_k - n_k/d - \ell}{n_k - \ell} \underset{k\to_\infty}{\rightarrow} 1 - \left(1 -  \frac{1}{d} \right)^d. \qed
\end{align*}

% \subsection{Proof of \Cref{prop:radThm} \label{app:proof_rad_lem}}

% \subsection{Proof of \Cref{lem:lipschitz} \label{app:proof_lipschitz}}

\subsection{Proof of \Cref{thm:stat_learning} \label{app:proof_stat_learn_thm}}

To establish \cref{thm:stat_learning}, we first define the Rademacher complexity of our sample as:
\begin{align} \label{eqn:rademacher}
\widehat{\mathcal{R}}:= \E_\sigma \left[ \sup_{x\in \CH(\X)} \frac{1}{K} \sum_{k=1}^K \sigma_k \left(1 - \frac{\OFF(x,D^k)}{Q}\right)\right],
\end{align}
where $(\sigma_k)_{k\in[K]}$ are independent Rademacher random variables, defined as random variables that are either $+1$ or $-1$, with probability $1/2$ each. We note that in~\eqref{eqn:rademacher} we are normalizing all values $\OFF(x,D^k)$ to lie in [0,1] (all rewards $r_{ij}$ lie in [0,1] and we have only $Q$ units, so the total reward is upper-bounded by $Q$) and subtracting it from 1 to get a loss function, to obtain the following result.
\begin{lemma}\label{prop:radThm}
\[ \E\left[ \sup_{x\in\CH(\X)} \hOFF(x)-\OFF(x)  \right] \leq 2Q\E_\wedge\left[\widehat{\mathcal{R}}\right]. \]
\end{lemma}
\proof{Proof.}  This inequality follows by applying a result shown within the proof of Theorem 3.3 in \citet{mohri2018foundations} for a general loss and a general family of functions to choose from. In our case, the loss corresponds to $1-\OFF(x,D)/Q$, and the family of functions to choose from is $\CH(\X)$. By plugging this in, we obtain
\[ \E\left[ \sup_{x\in\CH(\X)} \E_D\left[   1 - \frac{\OFF(x,D)}{Q}\right] - \frac{1}{K}\sum_{k=1}^K\left(1 - \frac{\OFF(x,D^k)}{Q}\right)\right] \leq 2\E_\wedge\left[\widehat{\mathcal{R}}\right].\]
The lemma follows by recognizing $\OFF(x) = \E_D[\OFF(x,D)]$, $\hOFF(x) = \sum_{k=1}^K \OFF(x,D^k)/K$, rearranging and multiplying both sides by $Q$.
\qed

The following two results that we directly quote will help us upper-bound the Rademacher complexity, thus giving an upper bound to the right-hand side in \Cref{prop:radThm}.
\begin{proposition}[Corollary 4 from \citet{maurer2016vector}]\label{prop:contraction}
        For functions $(h_k(x))_{k=1}^K$ that are $L$-Lipschitz on $x$ in the $\ell_2$ norm, we have
        \[ \E_{\sigma}\left[ \sup_{x\in \CH(\X)} \sum_{k=1}^K \sigma_k h_k(x) \right]\leq L\sqrt{2} \E_{\sigma'} \left[ \sup_{x\in \CH(\X)} \sum_{i=1}^n x_i \sum_{k=1}^K \sigma'_{ik}  \right], \]
        where $(\sigma_k)_{k\in[K]}$ and $(\sigma'_{ki})_{(k,i)\in[K]\times[n]}$ are independent Rademacher variables. 
    \end{proposition}

\begin{proposition}[Corollary D.11 from \citet{mohri2018foundations}]\label{prop:maxRandomWalk}
   \[ \E_{\sigma'}\left[ \max_{i\in [n]} \sum_{k=1}^K \sigma'_{ik}\right] \leq \sqrt{2K\log n}, \]
       where $(\sigma'_{ki})_{(k,i)\in[K]\times[n]}$ are independent Rademacher variables. 
   \end{proposition}

% With the above artillery, we just need to show to following before proving \Cref{thm:stat_learning}.
We now show that our objective function of interest is indeed Lipschitz in the placement $x$.

\begin{lemma}\label{lem:lipschitz}
    Functions $(h_k(x))_{k\in[K]}$ defined as $ h_k(x) = 1 -\OFF(x,D^k)/Q $ are $(\sqrt{n}/Q)$-Lipschitz in the $\ell_2$ norm.
\end{lemma}

\proof{Proof.} Consider two fractional inventory placements $x,x'\in \CH(\X)$. We have
\[ h_k(x) - h_k(x') = \frac{\OFF(x',D^k)-\OFF(x,D^k)}{Q}. \]
To bound $|\OFF(x',D^k)-\OFF(x,D^k)|$ notice that each if we increase only one component $x_i$ of $x$ by a quantity $\eta$, the objective value of the LP cannot increase by more than $\eta$ because the rewards $r_{ij}$ are upper bounded by 1. With this in mind, define $\eta_i = x_i - x_i'$ and write:
\[\OFF(x,D^k) - \OFF(x',D^k) = \sum_{i=1}^n \OFF\left(x+ \sum_{i'=1}^{i-1} \eta_{i'}e_{i'},D^k\right) - \OFF\left(x + \sum_{i'=1}^i \eta_{i'}e_{i'},D^k \right), \]
where $e_{i'}$ is a vector of zeros and a 1 in the $i'$-th component. Since $x+ \sum_{i'=1}^{i-1} \eta_{i'}e_{i'}$ and $x+ \sum_{i'=1}^{i} \eta_{i'}e_{i'}$ only differ in component $i$, we can use triangular inequality and bound
\begin{align*}
    |\OFF(x,D^k) - \OFF(x',D^k)| &\leq \sum_{i=1}^n \left|\OFF\left(x+ \sum_{i'=1}^{i-1} \eta_{i'}e_{i'},D^k\right) - \OFF\left(x + \sum_{i'=1}^i \eta_{i'}e_{i'},D^k \right) \right|\\
    &\leq \sum_{i=1}^n |\eta_i| = \sum_{i=1}^n |x_i - x_i'|= ||x-x'||_1 \leq \sqrt{n} || x-x'||_2. \qed
\end{align*}

\proof{Proof of \Cref{thm:stat_learning}.} To show the result we will upper bound the Rademacher complexity of a sample: $\widehat{\mathcal{R}}$. For this, we first combine \Cref{lem:lipschitz} with \Cref{prop:contraction} to bound
\[ \sup_{x\in \X}  \sum_{k=1}^K \sigma_k \left(1 - \frac{\OFF(x,D^k)}{Q}\right) \leq \frac{\sqrt{2n}}{Q} \E_{\sigma'} \left[ \sup_{x\in \CH(X)} \sum_{i=1}^n x_i \sum_{k=1}^K \sigma'_{ik}  \right]  = \frac{\sqrt{2n}}{Q}Q\E_{\sigma'}\left[\max_{i\in[n]}\sum_{k=1}^K \sigma'_{ik}\right], \]
where in the last equality we simply solve the supremum by placing all mass on the index $i$ with the highest $\sum_{k=1}^k \sigma'_{ik}$. Now, by \Cref{prop:maxRandomWalk}, the right-hand side of the above inequality is at most $\sqrt{2K\log n}$. Putting it all together, we get
\[ \widehat{\mathcal{R}} \leq\frac{1}{K} \sqrt{2n} \cdot \sqrt{2K\log n} = 2\sqrt{\frac{n\log n}{K}}. \]

To conclude we apply this upper bound to \Cref{prop:radThm} and obtain
\[ \E\left[ \sup_{x\in\X} \hOFF(x)-\OFF(x)  \right] \leq 2Q\E_\wedge\left[\widehat{\mathcal{R}}\right] \leq 4Q\sqrt{\frac{n\log n}{K}}.\qed\]

\subsection{Proof of \Cref{thm:main} \label{app:proof_main}}

We can lower-bound the left-hand side in \Cref{thm:main} as follows:
    \begin{align}
        \E_{\wedge,R}[\ONL(\pi^\alpha ,R(\hat{x})) ]& \geq \alpha \E_{\wedge,R}[\OFF(R(\hat{x}))] \label{eq:fulfillment_guarantee}\\
        & \geq \alpha \left(1 - \left(1 - \frac{1}{d}\right)^d\right) \E_\wedge\left[ \OFF(\hat{x})\right], \label{eq:beforeSampling}
    \end{align}
    where 
    in Inequality (\ref{eq:fulfillment_guarantee}) we used \Cref{assumpt:competitive} and in Inequality (\ref{eq:beforeSampling}) we used \Cref{thm:rand_rounding}. {Now, let $x^*$ denote an optimal solution to $\max_{x\in\CH(\X)}  \OFF(x)$,} and write
    \begin{align}
        \OFF(x^*) - \OFF(\hat{x}) = \left(\OFF(x^*) - \hOFF(x^*)\right) + \left(\hOFF(x^*) - \hOFF(\hat{x})\right) + \left(\hOFF(\hat{x}) - \OFF(\hat{x})\right).\label{eq:off_decomposition}
    \end{align}
    We know from {linearity of expectation} that $\E_\wedge[ \OFF(x^*) - \hOFF(x^*)] =0$ (note that $x^*$ is fixed and does not depend on the samples), and by definition $ \hOFF(x^*) - \hOFF(\hat{x})\leq 0$. For the last term in \Cref{eq:off_decomposition}, we use \Cref{thm:stat_learning} to bound
    \[ \E_\wedge\left[\hOFF(\hat{x}) - \OFF(\hat{x}) \right] \leq \E_\wedge\left[\sup_{x\in \CH(\X)}\hOFF(x) - \OFF(x) \right] =   O\left( Q\sqrt{\frac{n\log n}{K}}\right).  \]
       
       By taking expectations over $\wedge$ on \Cref{eq:off_decomposition} and rearranging we get
        \[  \E_\wedge[\OFF(\hat{x})] \geq  \OFF(x^*) - O\left( Q\sqrt{\frac{n\log n}{K}}\right) .\]
        Carrying on from Inequality (\ref{eq:beforeSampling}), we get
        \begin{align*}
            \alpha\left(1 - \left(1 - \frac{1}{d}\right)^d\right) \E_\wedge\left[ \OFF(\hat{x})\right] &\geq \alpha\left(1 - \left(1 - \frac{1}{d}\right)^d\right) \left( \OFF(x^*) - O\left( Q\sqrt{\frac{n \log n}{K}}\right)\right)\\
            & = \alpha\left(1 - \left(1 - \frac{1}{d}\right)^d\right)  \OFF(x^*) - O\left( Q\sqrt{\frac{n\log n}{K}}\right)\\
            & \geq \alpha\left(1 - \left(1 - \frac{1}{d}\right)^d\right)  \max_{x\in \X}\OFF(x) - O\left( Q\sqrt{\frac{n\log n}{K}}\right).\qed
        \end{align*}

\section{Greedy does not improve if $d=2$} \label{sec:greedySucks}

The sequence of examples in \citet[Thm.~3]{cornuejols1977exceptional} showing Greedy to be at best an $(1-(1-1/Q)^Q)$-approximation for any positive integer $Q$ can all be represented by an instance of our problem with $d=2$.

We first illustrate on the small case of $Q=3$.  There are 9 demand locations arranged in a grid.  The realized demand will be 1 at a uniformly chosen location, and 0 everywhere else.  There are 5 warehouses: one that serves each row, and one that serves each of the first two columns.  The rewards for serving any location in the first, second, and third columns are 3/9, 2/9, and 4/9 respectively, irrespective of the warehouse used.  The reward is 0 if a warehouse does not serve a location.
% Let $Q=3$.
Slightly perturbing rewards as necessary, Greedy would place a unit of inventory in the first column warehouse, followed by the second column warehouse, followed by a row warehouse.  It would cover 7 locations with rewards $\frac{3}{9},\frac{3}{9},\frac{3}{9},\frac{2}{9},\frac{2}{9},\frac{2}{9},\frac{4}{9}$ and hence earn expected reward $19/81$.
The optimal placement, on the other hand, is to place one unit of inventory in each row warehouse, covering all locations and earning expected reward $27/81$.
The approximation ratio is $19/27=1-(1-1/3)^3$, even though $d=2$ (each location is only served by its row warehouse and column warehouse).
% Such a transformation can be made for their example with any $Q$, and therefore Greedy is at best an $(1-1/e)$-approximation even if $d=2$.

This argument can be extended for general $Q$. Let there again be $Q^2$ demand locations arranged in a $Q$ by $Q$ grid, and the realized demand will be 1 at a uniformly chosen location and 0 everywhere else. There are $2Q-1$ warehouses: one for each row, serving all locations within the row, and one for each column except the last one, serving all locations within the column. We define the reward $r_i$ for all locations in column $i$ as follows, allowing them to exceed 1 without loss of generality (we can scale all rewards down by $Q^Q$ without changing the result.) For the last column, the reward is $r_Q = (Q-1)^{Q-1}$, and for the remaining columns the reward is defined recursively according to the following equations:
\begin{equation}
    Qr_i = \sum_{j=i}^Q r_j\quad \forall i\in[Q-1].\label{eq:grid_rewards}
\end{equation}
This equation can be interpreted as imposing that the sum of the rewards in column $i$ must be equal to the sum of the rewards in each row, summing from $i$ all the way to the right up to $Q$. It can be verified that $\sum_{i=1}^Q r_i = Q^{Q-1}$, so the sum of the rewards in the grid is equal to $Q^Q$.

If we slightly perturb the rewards so that the left hand side in \Cref{eq:grid_rewards} dominates, greedy will put a unit of inventory in every column and one row. This placement covers all locations except $Q-1$ of them in the last row, yielding an expected reward of
\[ \frac{Q^Q - (Q-1)r_Q}{Q^2} = \frac{Q^Q - (Q-1)^Q}{Q^2}. \]
On the other hand, an optimal placement places a unit in every row, yielding a reward of $Q^Q/Q^2$. Thus, the ratio between both is
\[   \frac{Q^Q - (Q-1)^Q}{Q^Q} = 1 - \left(1-\frac{1}{Q}\right)^Q.\]

\section{Incorporating Warehouse Capacity and Initial Inventories \label{app:capacity}}

As mentioned in \Cref{subsec:prob_statement}, all of our results directly extend to the setting where each warehouse has a capacity limiting the number of stock units that can be placed there, and where each warehouse also has a starting inventory. Let $C_i$ be the maximum number of units that can be stored at warehouse $i$, and let $S_i$ be the initial inventory at warehouse $i$ (that cannot be relocated). All of our results hold in this setting by solving the following generalization of $\max_{x\in \CH(\X)} \widehat{\OFF}(x)$:
\begin{align*}
    \max_{x\in \R^n_+, y\in\R^{[n]\times[m]\times[K]}_+} \quad & \frac{1}{K}\sum_{k=1}^K\sum_{i\in[n]}\sum_{j\in[m]} r_{ij}y_{ij}^k\\
    \mathrm{s.t.}\quad &\sum_{i=1}^n y_{ij}^k \leq D_j^k \quad &\forall j\in[m],k\in[K],\\
     &\sum_{i=j}^m y_{ij}^k \leq x_i + S_i \quad &\forall i\in[n],k\in[K],\\
     &x_i \leq C_i \quad &\forall i\in[n],\\
     & \sum_{i=1}^n x_i = Q.\\
\end{align*}
All of the proofs follow without change after this modification. We have to be careful in that, given the new constraints $x_i \leq C_i$ for all $i\in[n]$, if $Q$ is too big then constraint $\sum_{i=1}^n x_i = Q$ could make the problem infeasible. We can come around this by reducing $Q$ to the largest $Q'$ that makes the problem feasible. This change only lowers the number of samples required to obtain a sampling error of $\varepsilon$ in \Cref{thm:main}.

\section{Experiments}

\subsection{LP vs Greedy for optimizing Offline surrogate \label{app:lpvsgreedy}}

In \Cref{subsec:rounding} we mention that an alternative approach for optimizing the Offline surrogate is a greedy algorithm. We compare the solutions obtained through rounding the linear relaxation and the greedy algorithm by deploying them in our 90 instances. These are obtained with different combinations of networks (complete, long-chain, RDC-FDC), demand models ($\DHTI$, $\RHTI$, $\ROSI$), weights settings (uniform, rewards-based), and total inventory $Q\in \{30, 45, 60, 75, 90\}$.

For each of the 90 instances, we generate 500 samples of demand sequences and find solutions for $\max_{x\in\X}\hOFF(x)$ using both methods. Remarkably, all of the solutions found by the linear relaxation were integer, so no rounding was needed and the solutions were optimal. For the greedy procedure, we start with all warehouses empty. In each iteration, we solve the linear program definining $\hOFF(x)$ and place a unit of inventory in the warehouse whose inventory dual variable is the maximum. We iterate this procedure until we have placed all of the $Q$ units.

We find that both procedures produce very similar solutions in terms of placement and performance. For instance, the output of both procedures is exactly the same for 49 out of the 90 instances. For the instances where this is not the case, the output are still very alike. Out of the 41 instances with different placements, on average 2.8\% of the units are placed differently.

The solutions are hard to distinguish in terms of performance too. Since the optimal solution of the LP is integer, the reward obtained by the LP solution is always greater than the one obtained by the greedy procedure. That being said, across the 90 instances, the expected reward collected by the greedy solution is at least 99.75\% of the expected reward collected by the LP solution. To control for potential overfitting to the samples used to generate the solutions, we sampled 500 fresh demand samples and evaluated the expected rewards. We find that, in 90 instances, the average out-of-sample reward collected by the greedy solution is always within 99.63\% and 100.09\% of what the LP solution collects out-of-sample.

The only substantial difference that we found between both solutions was the time required to obtain each of them. For each instance, we look at the ratio between the runtime of the LP and the runtime of the greedy procedure. This ratio was at least 4 and at most 155, with an average value of 43. Given these findings, from this point on we will continue to use the LP relaxation and (potentially) rounding the solution to obtain the Offline placement.

\subsection{Description of Placement Procedures \label{app:placement_desc}}

\noindent\textbf{Fluid Placement.} This inventory placement solves the linear relaxation $\max_{x\in\CH{\X}} \FLU(x)$ and rounds the obtained solution using a greedy rounding scheme. The explicit linear relaxation that this inventory placement solves is the following:
\begin{align*}
\max_{x,y\geq0} \quad & \sum_{i\in[n]}\sum_{j\in[m]} y_{ij} r_{ij}\nonumber\\
\mathrm{s.t.}\quad & \sum_{i\in [n]} y_{ij} \leq \E[D_j] \quad &\forall j\in[m], \\
&\sum_{j\in[m]} y_{ij}\leq x_i \quad &\forall i\in[n],\\
&\sum_{i\in[n]} x_i = Q.
\end{align*}
In order to round a fractional solution, a greedy approach is taken. All fractional components are ordered from largest to smallest, and are rounded upwards one by one until the total inventory placement is exactly $Q$.

\noindent\textbf{Demand-scaled Fluid Placement.} This placement procedure again solves a fluid relaxation, but it modifies the expected demands so that they match the total inventory. Specifically, define $\rho = Q/\sum_{j\in [m]} \E[D_j]$ as the ratio between the total inventory and the expected total demand across all demand nodes. Then, the linear relaxation this placement procedure solves is
\begin{align*}
\max_{x,y\geq0} \quad & \sum_{i\in[n]}\sum_{j\in[m]} y_{ij} r_{ij}\nonumber\\
\mathrm{s.t.}\quad & \sum_{i\in [n]} y_{ij} \leq \rho \E[D_j] \quad &\forall j\in[m], \\
&\sum_{j\in[m]} y_{ij}\leq x_i \quad &\forall i\in[n],\\
&\sum_{i\in[n]} x_i = Q.
\end{align*}
Again, if a fractional solution is obtained, it will be rounded greedily.

\noindent\textbf{Offline Placement.} This placement procedure optimizes the sample-average approximation of the Offline Placement surrogate. Specifically, $K$ demand vectors are sampled (according to the corresponding demand model). Let $D^k_j$ be the demand originating from demand node $j$ on sample $k$. Then, the linear problem this surrogate solves is the following:
\begin{align*}
    \max_{x, y\geq 0} \quad & \frac{1}{K}\sum_{k=1}^K\sum_{i\in[n]}\sum_{j\in[m]} r_{ij}y_{ij}^k\\
    \mathrm{s.t.}\quad &\sum_{i=1}^n y_{ij}^k \leq D_j^k \quad &\forall j\in[m],k\in[K],\\
     &\sum_{i=j}^m y_{ij}^k \leq x_i \quad &\forall i\in[n],k\in[K],\\
     & \sum_{i=1}^n x_i = Q.\\
\end{align*}
As a remark, all of the solutions obtained by solving this linear program were integral, so no rounding was required.

\noindent\textbf{Myopic Placement.} This placement procedures uses the Myopic Surrogate. That is, it optimizes the inventory placement as if the fulfillment algorithm to be deployed downstream was myopic. This myopic algorithm chooses, out of the warehouses with remaining inventory, the one that offers the maximum immediate reward. This surrogate is optimized using a greedy procedure. Specifically, let $\MYO(x)$ denote the expected reward collected by the myopic fulfillment algorithm. The algorithm first initializes $x^{(0)} = (0)_{i\in{[n]}}$. Then, we iteratively set $x^{(\ell+1)} = \max_{i\in[n]} \MYO(x^{(\ell)} + e_i)$, where $e_i$ is a vector of zeros with a one on component $i$. The algorithm stops once $\sum_{i\in[n]} x^{(\ell)}_i = Q$, which is equivalent to stopping at $x^{(Q)}$. As with Offline Placement, this is solved through sample-average approximation, where $K$ demand sequences are sampled, and the value being optimized is the average reward collected by the myopic fulfillment policy across the $K$ samples.

\subsection{Description of Fulfillment Policies \label{app:fulfillment_desc}}

\noindent\textbf{Myopic Fulfillment.} In this policy, each demand node $j\in [m]$ has a fixed priority list over the warehouses $i\in [n]$, decreasing in $r_{ij}$. When demand from node $j$ arrives, the warehouse $i$ with the highest $r_{ij}$ and remaining inventory is chosen for fulfillment. 

\noindent\textbf{Fluid Shadow Prices.} This policy considers opportunity costs for depleting a unit of inventory in a certain warehouse through the shadow prices of the Fluid linear program. Before the online process begins, it solves the Fluid LP:
\begin{align}
\max_{y\geq0} \quad & \sum_{i\in[n]}\sum_{j\in[m]} y_{ij} r_{ij}\nonumber\\
\mathrm{s.t.}\quad & \sum_{i\in [n]} y_{ij} \leq \E[D_j] \quad &\forall j\in[m],\nonumber \\
&\sum_{j\in[m]} y_{ij}\leq x_i \quad &\forall i\in[n].\label{const:flu}
\end{align}
Let $\lambda_i$ be the dual variable for constraint \eqref{const:flu}. Then, whenever demand from node $j$ arrives, the warehouse $i$ that has the highest $r_{ij}-\lambda_i$ and remaining inventory is chosen for fulfillment.

\noindent\textbf{Offline Shadow Prices.} This policy considers opportunity costs for depleting a unit of inventory in a certain warehouse through the shadow prices of the Fluid linear program. Before the online process begins, it solves the Fluid LP:
\begin{align}
    \max_{x, y\geq 0} \quad & \frac{1}{K}\sum_{k=1}^K\sum_{i\in[n]}\sum_{j\in[m]} r_{ij}y_{ij}^k\nonumber\\
    \mathrm{s.t.}\quad &\sum_{i=1}^n y_{ij}^k \leq D_j^k \quad &\forall j\in[m],k\in[K],\nonumber\\
     &\sum_{i=j}^m y_{ij}^k \leq x_i \quad &\forall i\in[n],k\in[K].\label{const:off}
\end{align}
Let $\lambda_{ik}$ be the dual variable for constraint \eqref{const:off}. Then, whenever demand from node $j$ arrives, the warehouse $i$ with the highest $r_{ij}-\sum_{k=1}^K\lambda_{ik}$ and remaining inventory is chosen for fulfillment.

\noindent\textbf{Re-solving.} All of the above policies are static in the sense that they do not update their beliefs of the opportunity costs of depleting inventory units. This is suboptimal for two reasons. First, certain demand realizations can produce imbalanced inventory consumptions, producing scarcity in warehouses that started with abundant inventory, which should increase its opportunity cost. Second, information about demand arrivals during the time horizon can change our beliefs of future demand arrivals. For that, we also implement the re-solving versions of the Fluid and Offline shadow prices policies. In particular, we re-solve the corresponding LPs after the first arrival with arrival time of $\alpha \in \{1/3,2/3\}$, considering the current inventory levels and updating our beliefs about the demand distributions (except for the $\DHTI$ model).

\noindent\textbf{Posterior update.} The following Bayesian update is used to update the beliefs of the demand distributions. Let $T$ be a Geometric random variable with parameter $p$, representing the total number of arrivals, and let $X_1,\dots, X_T$ be the arrival times of each demand arrival (the order statistics of $T$ IID Uniform$[0,1]$ random variables). Let $T^\alpha$ be the number of random variables with arrival time at most $\alpha$. That is, $T_\alpha = |\{i:X_i <=\alpha\}|$. Let $T^{1-\alpha}$ be such that $T = T^\alpha + T^{1-\alpha}$. Then $T^{1-\alpha}|T^\alpha$ follows a negative binomial distribution with parameters $T^\alpha +1$ and $1-(1-p)(1-\alpha)$. This posterior is applied differently for the $\RHTI$ and $\ROSI$ models. In the former, we update our belief of the total number of demand arrivals, and they preserve the probabilities that they arrive from each demand node. In the latter, we update our beliefs for each demand node independently.

\newpage
\subsection{Results for Reward-based Weights \label{app:rw_weights}}

\begin{figure}[h]

    \centering
    \hspace*{-0.5cm}
    \begin{tikzpicture}
    % Begin a groupplot with shared legend options.
    \begin{groupplot}[
        group style={
            group size=3 by 3,
            vertical sep=1.2cm,
            horizontal sep=0.8cm,
            ylabels at=edge left,   % Place y-axis labels at the left edge
            xlabels at=edge bottom  % Place x-axis labels at the bottom edge
        },
        height=5cm, width=6cm,
        title style={yshift=1.2cm, font=\large},  % Shift column titles upward
        every axis y label/.style={
            at={(axis description cs:-0.3,0.5)},
            anchor=south,
            rotate=90
        },
        every axis title/.style={
            at={(axis description cs:0.5,1.2)},
            anchor=south
        },
        legend to name=globalLegend2,   % This collects legend entries under the name "globalLegend"
        legend columns=4               % Arrange the legend entries in one row
    ]
    
    % --- First Row of Plots ---
    % First subplot (the one where we add legend entries)
    \nextgroupplot[title={\shortstack{Deterministic-Horizon\\Temporal Independence}}, ylabel={Long-chain}, ymin=0.928, ymax=1.005]
        \addplot table [x=total_inventory, y=offline_lp_rounding, col sep=comma] {rewardbased_longchain_deter.csv};
        \addlegendentry{Offline LP rounding}
        \addplot table [x=total_inventory, y=fluid_lp_rounding, col sep=comma] {rewardbased_longchain_deter.csv};
        \addlegendentry{Fluid LP rounding}
        \addplot table [x=total_inventory, y=fluid_lp_rounding_withscaling, col sep=comma] {rewardbased_longchain_deter.csv};
        \addlegendentry{Fluid LP rounding with scaling}
        \addplot table [x=total_inventory, y=myopic_greedy, col sep=comma] {rewardbased_longchain_deter.csv};
        \addlegendentry{Myopic greedy}
    
    % Second subplot of the first row (no legend entries here)
    \nextgroupplot[title={\shortstack{Random-Horizon\\Temporal Independence}}, ymin=0.928, ymax=1.005, yticklabels={}]
        \addplot table [x=total_inventory, y=offline_lp_rounding, col sep=comma] {rewardbased_longchain_correl.csv};
        \addplot table [x=total_inventory, y=fluid_lp_rounding, col sep=comma] {rewardbased_longchain_correl.csv};
        \addplot table [x=total_inventory, y=fluid_lp_rounding_withscaling, col sep=comma] {rewardbased_longchain_correl.csv};
        \addplot table [x=total_inventory, y=myopic_greedy, col sep=comma] {rewardbased_longchain_correl.csv};
    
    % Third subplot of the first row
    
    \nextgroupplot[title={\shortstack{Random-Order\\Spatial Independence}}, ymin=0.928, ymax=1.005, yticklabels={}]
        \addplot table [x=total_inventory, y=offline_lp_rounding, col sep=comma] {rewardbased_longchain_indep.csv};
        \addplot table [x=total_inventory, y=fluid_lp_rounding, col sep=comma] {rewardbased_longchain_indep.csv};
        \addplot table [x=total_inventory, y=fluid_lp_rounding_withscaling, col sep=comma] {rewardbased_longchain_indep.csv};
        \addplot table [x=total_inventory, y=myopic_greedy, col sep=comma] {rewardbased_longchain_indep.csv};

    % --- Second Row of Plots ---
ch    \nextgroupplot[ylabel={\shortstack{RDC-FDC}}, ymin=0.765, ymax=1.005]
        \addplot table [x=total_inventory, y=offline_lp_rounding, col sep=comma] {rewardbased_starlike_deter.csv};
        \addplot table [x=total_inventory, y=fluid_lp_rounding, col sep=comma] {rewardbased_starlike_deter.csv};
        \addplot table [x=total_inventory, y=fluid_lp_rounding_withscaling, col sep=comma] {rewardbased_starlike_deter.csv};
        \addplot table [x=total_inventory, y=myopic_greedy, col sep=comma] {rewardbased_starlike_deter.csv};

    \nextgroupplot[ymin=0.765, ymax=1.005, yticklabels={}]
        \addplot table [x=total_inventory, y=offline_lp_rounding, col sep=comma] {rewardbased_starlike_correl.csv};
        \addplot table [x=total_inventory, y=fluid_lp_rounding, col sep=comma] {rewardbased_starlike_correl.csv};
        \addplot table [x=total_inventory, y=fluid_lp_rounding_withscaling, col sep=comma] {rewardbased_starlike_correl.csv};
        \addplot table [x=total_inventory, y=myopic_greedy, col sep=comma] {rewardbased_starlike_correl.csv};

    \nextgroupplot[ymin=0.765, ymax=1.005, yticklabels={}]
        \addplot table [x=total_inventory, y=offline_lp_rounding, col sep=comma] {rewardbased_starlike_indep.csv};
        \addplot table [x=total_inventory, y=fluid_lp_rounding, col sep=comma] {rewardbased_starlike_indep.csv};
        \addplot table [x=total_inventory, y=fluid_lp_rounding_withscaling, col sep=comma] {rewardbased_starlike_indep.csv};
        \addplot table [x=total_inventory, y=myopic_greedy, col sep=comma] {rewardbased_starlike_indep.csv};
    
    % --- Third Row of Plots ---
    \nextgroupplot[ylabel={Complete}, xlabel={Total inventory}, ymin=0.94, ymax=1.005]
        \addplot table [x=total_inventory, y=offline_lp_rounding, col sep=comma] {rewardbased_complete_deter.csv};
        \addplot table [x=total_inventory, y=fluid_lp_rounding, col sep=comma] {rewardbased_complete_deter.csv};
        \addplot table [x=total_inventory, y=fluid_lp_rounding_withscaling, col sep=comma] {rewardbased_complete_deter.csv};
        \addplot table [x=total_inventory, y=myopic_greedy, col sep=comma] {rewardbased_complete_deter.csv};

    \nextgroupplot[xlabel={Total inventory}, ymin=0.94, ymax=1.005, yticklabels={}]
        \addplot table [x=total_inventory, y=offline_lp_rounding, col sep=comma] {rewardbased_complete_correl.csv};
        \addplot table [x=total_inventory, y=fluid_lp_rounding, col sep=comma] {rewardbased_complete_correl.csv};
        \addplot table [x=total_inventory, y=fluid_lp_rounding_withscaling, col sep=comma] {rewardbased_complete_correl.csv};
        \addplot table [x=total_inventory, y=myopic_greedy, col sep=comma] {rewardbased_complete_correl.csv};

    \nextgroupplot[xlabel={Total inventory}, ymin=0.94, ymax=1.005, yticklabels={}]
        \addplot table [x=total_inventory, y=offline_lp_rounding, col sep=comma] {rewardbased_complete_indep.csv};
        \addplot table [x=total_inventory, y=fluid_lp_rounding, col sep=comma] {rewardbased_complete_indep.csv};
        \addplot table [x=total_inventory, y=fluid_lp_rounding_withscaling, col sep=comma] {rewardbased_complete_indep.csv};
        \addplot table [x=total_inventory, y=myopic_greedy, col sep=comma] {rewardbased_complete_indep.csv};
    
    \end{groupplot}
    \end{tikzpicture}
    
    % Now place the shared legend (adjust its placement as desired)

    % Now place the shared legend by referencing its name
    \vspace{0.5cm} % Adjust spacing as needed
  \begin{tikzpicture}[scale=1]
  % Offline LP rounding (blue, with a circle marker)
  \draw[myBlue, thick, dashed,
    decoration={
      markings,
      mark=at position 0.5 with {
        \node[fill=myBlue, draw=myBlue, solid, shape=circle, inner sep=2pt] {};
      }
    },
    postaction={decorate}
  ] (0,0) -- (1,0);
  \node[right] at (1.2,0) {Offline Placement};

  % Fluid LP rounding (orange, with a solid square marker)
  \draw[myOrange, thick, dashed,
    decoration={
      markings,
      mark=at position 0.5 with {
        \node[fill=myOrange, draw=myOrange, solid, shape=rectangle, inner sep=2.5pt] {};
      }
    },
    postaction={decorate}
  ] (0,-0.5) -- (1,-0.5);
  \node[right] at (1.2,-0.5) {Fluid Placement};

  % Fluid LP rounding with scaling (green, with a triangle marker)
  \draw[myGreen, thick, dashed,
    decoration={
      markings,
      mark=at position 0.5 with {
        \node[fill=myGreen, draw=myGreen, solid, 
          shape=regular polygon, regular polygon sides=3, inner sep=1.3pt, rotate=120] {};
      }
    },
    postaction={decorate}
  ] (0,-1) -- (1,-1);
  \node[right] at (1.2,-1) {Scaled Demand Fluid Placement};

  % Myopic greedy (magenta, with a diamond marker)
  \draw[myMagenta, thick, dashed,
    decoration={
      markings,
      mark=at position 0.5 with {
        \node[fill=myMagenta, draw=myMagenta, solid, shape=diamond, inner sep=1.8pt] {};
      }
    },
    postaction={decorate}
  ] (0,-1.5) -- (1,-1.5);
  \node[right] at (1.2,-1.5) {Myopic Placement};
\end{tikzpicture}

\caption{Competitive ratio as a function of total inventory for all combinations of graphs and demand models with reward-based weights. Each row corresponds to a different network structure and each column corresponds to a different demand model.}
\label{fig:competitiveratiorwb}
\end{figure}

\end{APPENDICES}

\end{document}